\documentclass[aps,prc,preprint,groupedaddress,showpacs]{revtex4-1}
\usepackage{natbib}
\usepackage{graphicx}
\usepackage{amsmath}
\usepackage{hyperref}
\usepackage{color}
\allowdisplaybreaks

\begin{document}

\preprint{JLAB-THY-15-2114}

\title{Coincidence charged-current neutrino-induced deuteron disintegration}

\author{O. Moreno}
\author{T. W. Donnelly}
\affiliation{Center for Theoretical Physics, Laboratory for Nuclear Science and Department of Physics, Massachusetts Institute of Technology, Cambridge, MA 02139, USA}
\author{J. W. Van Orden}
\affiliation{Department of Physics, Old Dominion University, Norfolk, VA 23529\\ Jefferson Lab,12000 Jefferson Avenue, Newport News, VA 23606, USA\footnote{Notice: Authored by Jefferson Science Associates, LLC under U.S. DOE Contract No. DE-AC05-06OR23177.
The U.S. Government retains a non-exclusive, paid-up, irrevocable, world-wide license to publish or reproduce this manuscript for U.S. Government purposes.}
}
\author{W. P. Ford}
\affiliation{Department of Nuclear Engineering,
University of Tennessee,
315 Pasqua Nuclear Engineering,
Knoxville, TN 37916
}


\begin{abstract}
Deuteron disintegration by charged-current neutrino (CC$\nu$) scattering offers the possibility to determine the energy of the incident neutrino by measuring in coincidence two of the three resulting particles: a charged lepton (usually a muon) and two protons, where we show that this channel can be isolated from all other, for instance, from those with a pion in the final state. We discuss the kinematics of the process for several detection scenarios, both in terms of kinematic variables that are natural from a theoretical point of view and others that are better matched to experimental situations. The deuteron structure is obtained from a relativistic model (involving an approximation to the Bethe-Salpeter equation) as an extension of a previous, well-tested model used in deuteron electrodisintegration. We provide inclusive and coincidence (semi-inclusive) cross sections for a variety of kinematic conditions, using the plane-wave impulse approximation, introducing final-state hadronic exchange terms (plane-wave Born approximation) and final-state hadronic interactions (distorted-wave Born approximation).

\end{abstract}

\pacs{25.30.Pt, 12.15.Ji, 13.15.+g, 21.45.Bc}

\maketitle

\section{Introduction \label{introduction}}

Deuterium has been used in the past as a target in several neutrino quasielastic scattering experiments for average energies in the range 0.5 -- 27 GeV, where  the final-state particles in the reaction were detected in bubble chambers \cite{Gallagher:2011zza}. For reactions below pion production threshold, a neutrino interacts with a neutron bound in deuteron, which turns into a proton and a negatively-charged lepton is produced. The two resulting protons are no longer bound and one or both of them can be detected in coincidence with the charged lepton.

From a theoretical point of view, the advantage of using deuterium as a target with coincidence detection lies in the fact that the kinematics of the remaining nucleon, and of all of the particles taking part in the process, are fully determined by energy-momentum conservation. This includes the energy of the incoming neutrino, which usually has a broad energy spectrum, an issue for neutrino oscillation experiments when nuclei other than deuterium are used. In the case of nuclei in general, the nucleus is in general excited to configurations at high missing energy (see \cite{semi} for a discussion of semi-inclusive CC$\nu$ reactions including the definition of missing energy) and, even when both a charged lepton and a proton are detected in coincidence, one cannot reconstruct the incident neutrino energy. However, the deuteron does not share this problem and all kinematic variables can be reconstructed from measurements of a subset of final-state particles. The details of such procedures are discussed later. We note that, insofar as a cut can be made to separate events where pions are produced from those where they cannot, {\it i.e.,} where the ``no-pion" cross section can be isolated (the kinematics for making such a cut are discussed below), what we continue to call the semi-inclusive cross section is actually an exclusive cross section, meaning that the energies and momenta of all particles are determined by measuring only the subsets summarized above.

In addition to determining the neutrino energy using only the kinematics of the reaction being studied, namely, $\nu_{\mu} + {}^2$H$\rightarrow \mu^- + p + p$, the cross section for this reaction can be used to determine the neutrino flux. This, of course, requires that one knows that cross section, and indeed, especially under favorable conditions, this is the case as discussed later. Such is not the case, however, for complex nuclei such as carbon or oxygen where considerable effort has gone into evaluating the level of theoretical uncertainty in modeling neutrino reactions in those cases. 

Aside from providing a means to determine the energy and flux of the incident neutrino in such semi-inclusive CC$\nu$ reactions, neutrino disintegration of the deuteron has the potential to yield valuable new information on the nucleonic content in the problem. Specifically, given that the vector EM form factors of the nucleon are relatively well-determined from electron scattering on protons and light nuclei, the prime candidate for such an approach is the isovector, axial-vector form factor of the nucleon, $G_A^{(1)}$. Once the uncertainties in the modeling of the deuteron and $pp$ systems have been evaluated, {\it i.e.,} the extent to which different ground-state wave functions yield different results and to which the treatment of the $pp$ final state gives different answers, the reaction can be used to determine $G_A^{(1)}$. Both aspects of the problem are discussed below.

The paper is organized as follows: In Sect. \ref{formalism_kinematics} the basic kinematics of the process is described, and the formalism of the weak responses and the neutrino-deuteron cross section is given in Sect. \ref{formalism_crosssection}.  In Sect. \ref{formalism_deuteron} the deuteron structure model is summarized. In Sect. \ref{results} results are presented and discussed for inclusive and semi-inclusive neutrino-deuteron scattering for several choices of kinematics. Finally, in Sect. \ref{conclusions} our conclusions are given.

\section{Kinematics \label{formalism_kinematics}} 

\begin{figure}
\begin{center}
\includegraphics[width=0.8\textwidth]{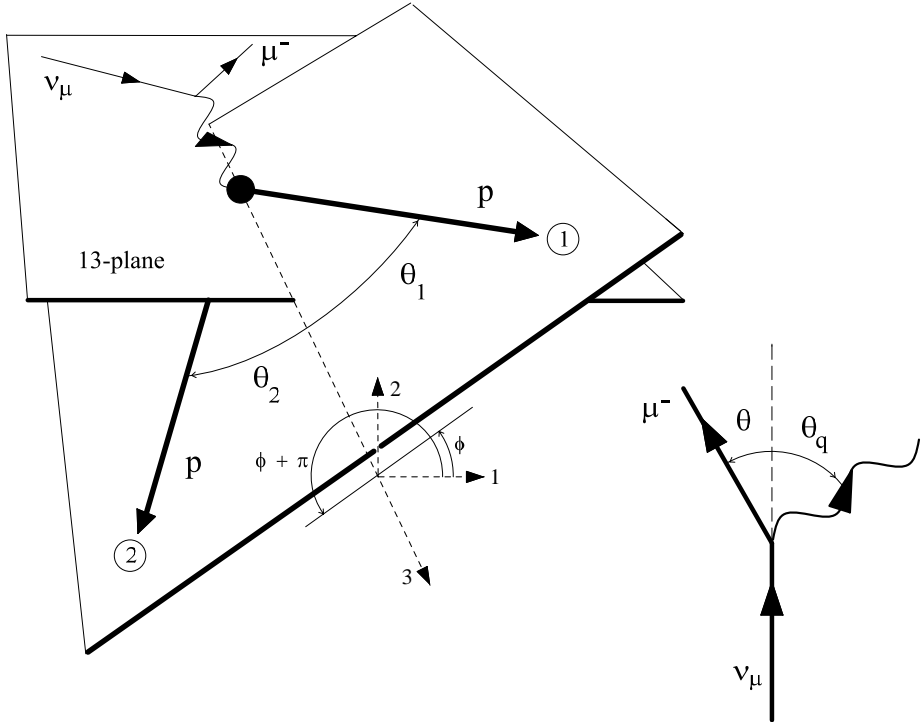}
\caption{Kinematics for coincidence neutrino-deuteron reactions. \label{fig_kinematics}}
\end{center}
\end{figure}

Let us begin from a ``theoretical'' point of view and presume that the energy of the incident neutrino is known, having mass $m$ with three-momentum $\mathbf{k}$ and total energy $\varepsilon =\sqrt{k^{2}+m^{2}}$, contained in the four-momentum $K^{\mu }=(\varepsilon ,\mathbf{k})$. Later we will work backwards, assuming only final-state momenta are known and show how the neutrino energy may be reconstructed. The four-momentum corresponding to the outgoing charged lepton, with mass $m^{\prime}$, is $K^{\prime \mu }=(\varepsilon^{\prime },\mathbf{k}^{\prime })$, and the four-momentum transfer is $Q^{\mu }=(\omega ,\mathbf{q})$, with $-Q^2=|Q^2|=q^2-\omega^2\ge 0$ (spacelike). The three-momentum transfer $\mathbf{q}$ is assumed to be along the $3$-axis so that the lepton momenta define the $13$-plane (see Fig. \ref{fig_kinematics}), and the angle between them is the scattering angle $\theta$. With these definitions the components of the above mentioned four-momenta can be written as
\begin{equation}
\begin{array}{lll}
K^{0} = \varepsilon & K^{\prime 0} = \varepsilon^{\prime}  & Q^{0} =\varepsilon -\varepsilon^{\prime} = \omega \\
K^{1} = \frac{1}{q}kk^{\prime }\sin \theta & K^{\prime 1} = \frac{1}{q}kk^{\prime }\sin \theta & Q^{1} = 0 \\
K^{2} = 0 & K^{\prime 2} = 0 & Q^{2} = 0 \\
K^{3} = \frac{1}{q}k\left( k-k^{\prime }\cos \theta \right) \quad\quad & K^{\prime 3} = -\frac{1}{q}k^{\prime} \left(k^{\prime}-k\cos \theta \right) \quad\quad & Q^{3} =\sqrt{k^{2}+k^{\prime 2}-2kk^{\prime }\cos \theta} = q \;.
\end{array}
\label{lepton_momenta}
\end{equation}

Concerning the hadronic part of the process, the deuteron at rest carries four-momentum $P_{d}^{\mu}=(M_{d},0,0,0)$, where $M_{d}= $1875.61 MeV is its mass. The protons resulting from the CC$\nu$ disintegration of the target have four-momenta $P_{1}^{\mu }=(E_{1},\mathbf{p}_{1})$ and $P_{2}^{\mu }=(E_{2},\mathbf{p}_{2})$ respectively.
Let $\theta^{\prime}_{1}$ and $\theta^{\prime}_{2}$ be, respectively, the angles of the proton momenta with respect to the incident neutrino momentum and $\phi^{\prime}_{1}$ and $\phi^{\prime}_{2}$ their angles with respect to the leptonic plane; the components of the proton four-momenta can then be written as
\begin{equation}
\begin{array}{ll}
P_{1}^{0}=E_{1} & P_{2}^{0}=E_{2} \\
P_{1}^{1} = p_{1} \left[ \sin\theta^{\prime}_{1} \cos\phi^{\prime}_{1} \cos\theta_q + \cos\theta^{\prime}_{1} \sin\theta_q \right]  & P_{2}^{1} = p_{2} \left[ \sin\theta^{\prime}_{2} \cos\phi^{\prime}_{2} \cos\theta_q + \cos\theta^{\prime}_{2} \sin\theta_q \right] \\
P_{1}^{2} = p_{1} \sin\theta^{\prime}_{1} \sin\phi^{\prime}_{1} & P_{2}^{2} = p_{2} \sin\theta^{\prime}_{2} \sin\phi^{\prime}_{2} \\ 
P_{1}^{3} = p_{1} \left[ -\sin\theta^{\prime}_{1} \cos\phi^{\prime}_{1} \sin\theta_q  + \cos\theta^{\prime}_{1} \cos\theta_q \right] \qquad & P_{2}^{3} = p_{2} \left[ -\sin\theta^{\prime}_{2} \cos\phi^{\prime}_{2} \sin\theta_q  + \cos\theta^{\prime}_{2} \cos\theta_q \right]\;,
\end{array}
\label{protons_momenta}
\end{equation}
where $\theta_q$ is the angle between the incident momentum $\mathbf{k}$ and the momentum transfer $\mathbf{q}$, so that
\begin{equation}
\sin\theta_q = \frac{1}{q} k^{\prime} \sin\theta \qquad \text{and} \qquad  \cos\theta_q = \frac{1}{q} (k-k^{\prime} \cos\theta)\;.
\end{equation}

The primed angles $\theta^{\prime}_1$, $\theta^{\prime}_2$, and $\phi^{\prime}_{1}, \phi^{\prime}_{2}$, defined with respect to the incident neutrino momentum, are related to the unprimed ones defined with respect to the momentum transfer, as were used in \cite{semi}, through
\begin{eqnarray}\nonumber
&& \sin\theta^{\prime}_i \cos\phi^{\prime}_i = \sin\theta_i \cos\phi_i \cos\theta_q - \cos\theta_i \sin\theta_q \\
&& \sin\theta^{\prime}_i \sin\phi^{\prime}_i = \sin\theta_i \sin\phi_i \\\nonumber
&& \cos\theta^{\prime}_i = \sin\theta_i \cos\phi_i \sin\theta_q + \cos\theta_i \cos\theta_q
\end{eqnarray}
with $\phi_{2}=\phi_{1}+\pi$. Let us next turn to the reverse situation, the ``experimental'' point of view where only final-state particles are presumed to be detected and where the goal is to determine the incident neutrino energy.

\subsection{Neutrino energy determination}

\begin{figure}
\begin{minipage}[p]{0.48\linewidth}
\centering
\includegraphics[width=1.1\textwidth] {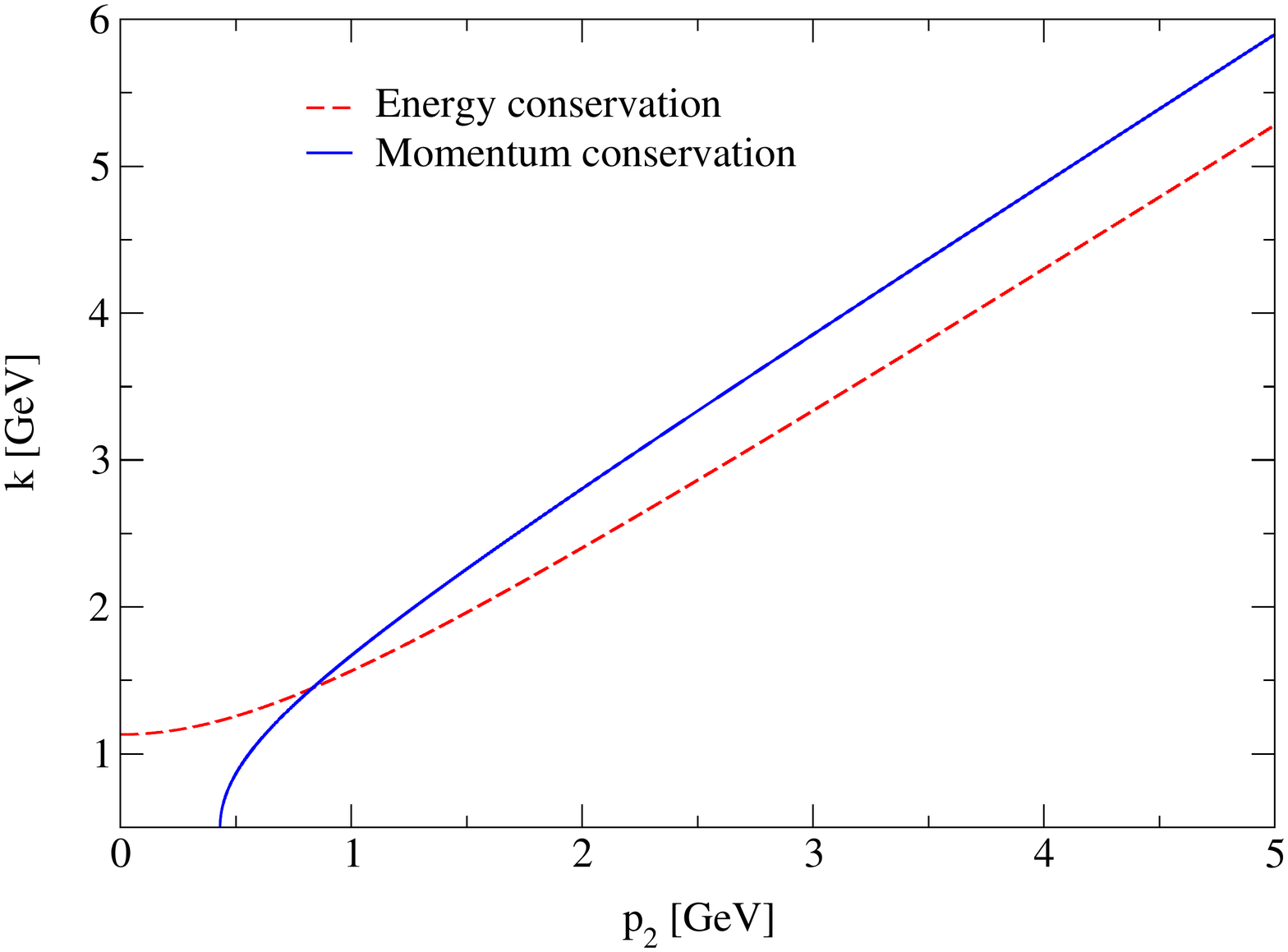}
\end{minipage}
\hspace{0.15in}
\begin{minipage}[p]{0.48\linewidth}
\centering
\includegraphics[width=1.1\textwidth] {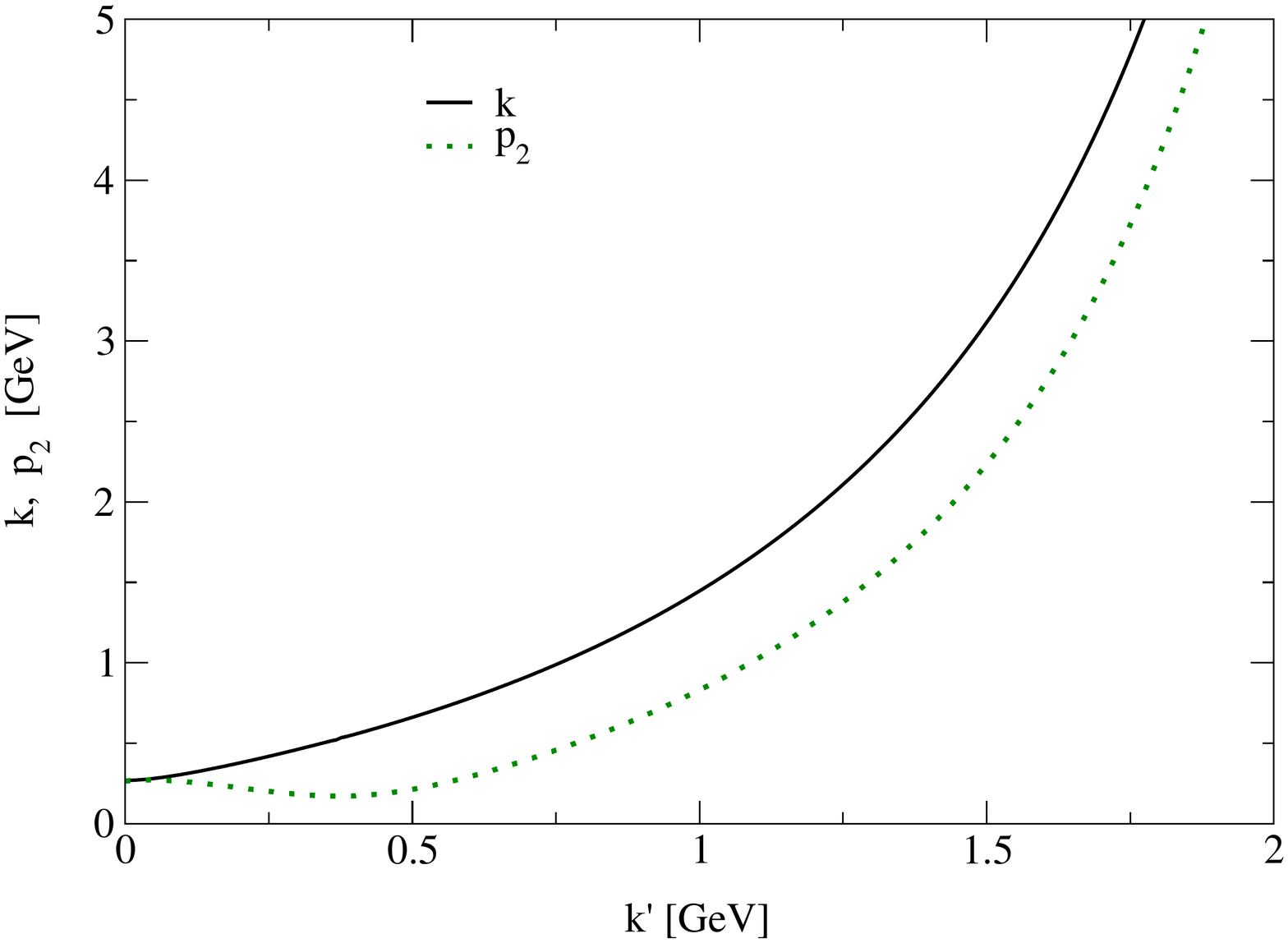}
\end{minipage}
\caption{(Color online) Left: An example of using energy conservation (dashed curve) and momentum conservation (solid curve), to determine the incident neutrino momentum $k$ as a function of the undetected proton momentum $p_2$ for a deuteron disintegration with emitted muon momentum $k^{\prime}=$ 1 GeV, scattering angle $\theta=$ 60$^{\circ}$, emitted proton momentum $p_1=$ 0.5 GeV, and proton emission angle $\theta_1=$ 20$^{\circ}$. The intersection of the curves gives the only possible combination of incoming neutrino and undetected proton momenta for this process (1.45 GeV and 0.83 GeV, respectively). Right: Using this procedure, and for the same kinematic conditions for $\theta$, $p_1$ and $\theta_1$, resulting incident neutrino momentum $k$ (solid curve) and undetected proton momentum $p_2$ (dotted curve) as a function of the emitted muon momentum $k'$. \label{fig_neutrino_determination}}
\end{figure}

On the one hand, energy and momentum conservation in the hadronic vertex implies $\omega =E_{1}+E_{2}-M_{d}$ and $\mathbf{q}=\mathbf{p}_{1}+\mathbf{p}_{2}$, or $q = p_1\cos\theta_1 \pm \sqrt{p_2^2 - p_1^2\sin^2\theta_1}$ for its magnitude. On the other hand, from the lepton vertex one has, as shown in Eqs.~(\ref{lepton_momenta}), $\omega= \varepsilon -\varepsilon^{\prime}$ and $q=\sqrt{k^{2}+k^{\prime 2}-2kk^{\prime }\cos \theta}$. The transfer variables $\omega$ and $q$ must be equal in both vertices, and positive. Hence, the energy-momentum conservation conditions yield two relationships between the undetected proton energy $E_2$, or corresponding momentum $p_2$, and the incident neutrino energy $\varepsilon$, or corresponding momentum $k$; the only possible values of $E_2$ and $\varepsilon$ (or $p_2$ and $k$) for the process are those that fulfill both conditions simultaneously. It should be noted that this procedure cannot be followed with a general nuclear target, since the relationship between $E_2$ and $p_2$ is not fixed. Indeed, although on-shellness always holds, namely $E_{A-1}^2=p^2_{A-1}+W^2_{A-1}$, the rest mass of the residual system $W_{A-1}$ is in general unknown (see \cite{semi}); however, in the specific case of the deuteron target, $W_{A-1}=m_p$ without ambiguity. Fig. \ref{fig_neutrino_determination} shows a graphical example of how the constraints work. In the following subsection we develop this possibility of neutrino energy determination in deuteron scattering analytically by defining and studying specific kinematic scenarios, as well as the pion production threshold and other kinematic constraints.

\subsection{Kinematic scenarios} \label{kinematic_scenarios}

We consider two scenarios, the first (`scenario A') where the two protons in
the final state are presumed to be detected, but not the muon, and the
second (`scenario B') where the final-state charged lepton and one proton
(called proton number 1) are presumed to be detected, but not the other
proton (proton number 2). Using the nomenclature given above, we define the
following four-momenta:
\begin{eqnarray}
P_{A}^{\mu }=(E_{A},\boldsymbol{p}_{A}) &\equiv & P_{1}^{\mu }+P_{2}^{\mu }
\label{eq-k1} \\
P_{B}^{\mu }=(E_{B},\boldsymbol{p}_{B}) &\equiv & K^{\prime \mu }+P_{1}^{\mu } \;,  \label{eq-k2}
\end{eqnarray}%
which yield the following relationships:%
\begin{eqnarray}
E_{A} &=&E_{1}+E_{2}=M_{d}+\omega   \label{eq-k3} \\
\boldsymbol{p}_{A} &=&\boldsymbol{p}_{1}+\boldsymbol{p}_{2}=\boldsymbol{q}
\label{eq-k4} \\
E_{B} &=&\varepsilon ^{\prime }+E_{1}  \label{eq-k5} \\
\boldsymbol{p}_{B} &=&\boldsymbol{k}^{\prime }+\boldsymbol{p}_{1}\;.
\label{eq-k6}
\end{eqnarray}%
Note that in scenario A this means that $q$ and $\omega $ are immediately
known. The angles between the incoming neutrino direction and the three-vectors $%
\boldsymbol{p}_{A}$ and $\boldsymbol{p}_{B}$ are known and denoted $\theta
_{A}$ and $\theta _{B}$, respectively. Referring to Fig. \ref{fig_kinematics} we see that $%
\theta _{A}=\theta _{q}$. Let us now develop the two scenarios one at a time.\\

\textbf{Scenario A:}
In this case one knows $\boldsymbol{q}$ and $\omega$ through $\boldsymbol{q}=\boldsymbol{p}_A$ and $\omega=E_A-M_d$. Noting that $\boldsymbol{k}^{\prime }=\boldsymbol{k}-\boldsymbol{q}$, one has
\begin{eqnarray}
k^{\prime 2} &=&k^{2}+q^{2}-2kq\cos \theta _{q}  \label{eq-k9}
\end{eqnarray}%
and together with $\varepsilon ^{\prime}=\varepsilon -\omega$, and using the fact that $\varepsilon ^{\prime 2}-k^{\prime 2}=m^{\prime 2}$, one obtains
\begin{equation}
k\:q\cos \theta _{q}-\omega \:\varepsilon = X_{A}  \label{eq-k11}
\end{equation}%
with
\begin{equation}
X_A \equiv \frac{1}{2}\left[ q^2-\omega^2+m^{\prime
2}-m^{2}\right] >0\;.  \label{eq-k11}
\end{equation}%
This immediately leads to values of the incoming neutrino momentum and
energy:%
\begin{eqnarray}
k &=&\frac{1}{a_{A}}\left[ q\cos \theta _{q}X_{A}+\omega \sqrt{%
X_{A}^{2}+m^{2}a_{A}}\right]   \label{eq-k12} \\
\varepsilon  &=&\frac{1}{a_{A}}\left[ \omega X_{A}+q\cos \theta _{q}\sqrt{%
X_{A}^{2}+m^{2}a_{A}}\right] \;,  \label{eq-k13}
\end{eqnarray}%
where%
\begin{equation}
a_{A}\equiv \left( q\cos \theta _{q}\right) ^{2}-\omega ^{2}\geq 0\;.
\label{eq-k14}
\end{equation}%
One can show that $0\leq \theta _{q}\leq \theta _{q}^{0}\leq \pi /2$, where $%
\theta _{q}^{0}\equiv \arccos (\omega /q)$. Knowing the neutrino beam energy 
$\varepsilon $ then yields the charged lepton energy $\varepsilon ^{\prime
}=\varepsilon -\omega $ and thus its momentum $k^{\prime }=\sqrt{\varepsilon
^{\prime 2}-m^{\prime 2}}$, together with the angle $\theta $ through
\begin{eqnarray}
k^{\prime }\cos \theta  &=&k-q\cos \theta _{q}  \label{eq-k7} \\
k^{\prime }\sin \theta  &=&q\sin \theta _{q}\;.  \label{eq-k8}
\end{eqnarray}%
That is, one now has all of the kinematic variables in scenario A.\\

\textbf{Scenario B:}
In this situation $E_{B}$ and $\boldsymbol{p}_{B}$ are presumed to be known, and the unknown proton variables can be expressed as $E_{2} =\varepsilon - (E_B-M_d)$ and $\boldsymbol{p}_{2} =\boldsymbol{k}-\boldsymbol{p}_{B}$; from the latter one gets
\begin{eqnarray}
p_2^2 &=& k^{2}+p_B^{2}-2\:k\:p_B\cos \theta _{B}  \;,\label{eq-k9}
\end{eqnarray}
and then using the fact that $E_{2}^{2}-p_{2}^{2}=m_{p}^{2}$ one has
\begin{equation}
k\:p_{B}\:\cos \theta_{B} - (E_B-M_d)\:\varepsilon=  X_{B} 
  \label{eq-k18}
\end{equation}
with
\begin{equation}
X_{B} \equiv \frac{1}{2}\left[p_{B}^{2}-(E_B-M_d)^{2}+m_{p}^{2}-m^{2}\right]  > 0
  \label{eq-k18}
\end{equation}
from which one obtains%
\begin{eqnarray}
k &=&\frac{1}{a_{B}}\left[ p_{B}\cos \theta_{B}X_{B}+(E_B-M_d) \sqrt{X_{B}^2+m^{2}a_{B}}\right] 
\label{eq-k20} \\
\varepsilon  &=&\frac{1}{a_{B}}\left[(E_B-M_d)\:X_{B}+p_{B}\cos\theta _{B} \sqrt{X_{B}^2+m^{2}a_{B}}\right] \;,
\label{eq-k21}
\end{eqnarray}
where 
\begin{equation}
a_{B}\equiv \left( p_{B}\cos \theta _{B}\right)^{2}-(E_B-M_d)^{2}\;.  \label{eq-k22}
\end{equation}%
In this case, one now has $q$ and $\omega $ using the usual relationships in Eqs. (\ref{lepton_momenta}) and can then find $E_{2}$ and $\boldsymbol{p}_{2}$ using Eqs. (\ref{eq-k3}) and (\ref{eq-k4}), yielding all of the kinematic variables in scenario B.\\

We now study the incident neutrino energy or momentum threshold to produce the lightest possible extra particle in the scattering process, namely a neutral pion, $\pi^0$. One assumes the knowledge of the three emitted particles variables, and then defines
\begin{equation}
{P}_{C}^{\mu } =(E_C,\boldsymbol{p}_C) \equiv K'^{\mu }+P_{1}^{\mu }+P_{2}^{\mu } \;, \label{pc}
\end{equation}%
so that $E_{C}=\varepsilon'+E_1+E_2$ and $\boldsymbol{p}_{C}=\boldsymbol{k'}+\boldsymbol{p}_{1}+\boldsymbol{p}_{2}$. One then has $E_{\pi^0}=M_d+\varepsilon-E_C$ and $\boldsymbol{p}_{\pi^0}=\boldsymbol{k}-\boldsymbol{p}_C$, whence
\begin{eqnarray}
p_{\pi^0}^2 &=& k^{2}+p_C^{2}-2\:k\:p_C\cos \theta _{C}  \;,\label{p_pi}
\end{eqnarray}
and using the on-shell relation for an emitted neutral pion, $E_{\pi^0}^2-p_{\pi^0}^2=m_{\pi^0}^2$, one obtains
\begin{equation}
k\:p_{C}\:\cos\theta_{C} - (E_C-M_d)\:\varepsilon=  X_{C} \;,
\end{equation}
with
\begin{equation}
X_C \equiv  \frac{1}{2}\left[p_{C}^{2}-(E_C-M_d)^{2}+m_{\pi^0}^{2}-m^{2}\right] \;.
  \label{XC}
\end{equation}
From these expressions one obtains the following pion production threshold values of the incident neutrino momentum and energy:
\begin{eqnarray}
k_{\pi^0\: th.} &=&\frac{1}{a_{C}}\left[p_{C}\cos \theta_{C} \:X_{C} \pm (E_C-M_d) \sqrt{X_{C}^2+m^{2}a_{C}}\right] 
\label{k_pion} \\
\varepsilon_{\pi^0\: th.}   &=&\frac{1}{a_{C}}\left[(E_C-M_d) \:X_{C} \pm p_{C}\cos\theta _{C} \sqrt{X_{C}^2+m^{2}a_{C}}\right] \;,
\label{e_pion}
\end{eqnarray}
where
\begin{equation}
a_{C}\equiv \left( p_{C}\cos \theta _{C}\right)^{2}-(E_C-M_d)^{2}\;.  \label{aC}
\end{equation}
For given energies of the emitted particles $\varepsilon'$, $E_1$, $E_2$, neutral pion production is ruled out in the scattering process if the incident neutrino beam energy fulfills $\varepsilon\le \varepsilon_{\pi^0\:th.}$ (and equivalently for the momenta).
\\
A potential strategy for measurements of the desired kinematic variables might be the following: one might assume that the three particles in the final state, say, a muon and the two protons, are all measured with adequate precision first to eliminate the possibility of $\pi^0$ production, or, more generally, to isolate processes where a pion is produced from those where it is not, since the former is interesting in its own right. Given this first cut, one can then safely proceed to analyze the reaction
\begin{equation}
\nu_{\mu} + \:^2\text{H} \to \mu^- + p + p \nonumber
\end{equation}
as above in either of the scenarios, whichever proves to be the more favorable from an experimental point of view.
\\
Finally in this section, for completeness 
it is also useful to make contact with the general developments of
semi-inclusive CC$\nu$ reactions presented in \cite{semi}.
The variables introduced there translate into the present ones in the
following way:%
\begin{eqnarray}
M_{A}^{0} \leftrightarrow M_{d} \;;\qquad m_{N} \leftrightarrow m_{p}  \;;\qquad 
W_{A-1} \leftrightarrow m_{p}  \;;\qquad \boldsymbol{p}_{N} \leftrightarrow \boldsymbol{p}_{1} \;;\qquad \boldsymbol{p} \leftrightarrow -\boldsymbol{p}_{2} \;.
\label{eq-k23}
\end{eqnarray}%
The momentum of what is labeled particle 2 above is usually called the missing momentum: $\boldsymbol{p}_{m}=\boldsymbol{p}_{2}=-\boldsymbol{p}$. For simplicity below we will show results as functions of $p=|\boldsymbol{p}|$. The quantity in that paper called $\mathcal{E}$ in previous scaling analyses
(see \cite{Day:1990mf,Donnelly:1998xg,Donnelly:1999sw}) is zero, since the daughter
system in the general case is simply the second proton in the present study.
This leads to the following for the scaling variable $y$ and the quantity $Y$:
\begin{eqnarray}
y &=&\frac{1}{2}\left[ \frac{(M_{d}+\omega )}{W}\sqrt{W^{2}-W_{T}^{2}}-q
\right]   \label{eq-k24} \\
Y &=&q+y\;,  \label{eq-k25}
\end{eqnarray}%
where the invariant mass in the final state is given by%
\begin{equation}
W=\sqrt{\left( M_{d}+\omega \right) ^{2}-q^{2}}  \label{eq-k26}
\end{equation}%
and its threshold value by $W_{T}=2m_{p}$, so that $W\geq W_{T}$.
A useful relationship that emerges is%
\begin{equation}
M_{d}+\omega =\sqrt{m_{p}^{2}+y^{2}}+\sqrt{m_{p}^{2}+Y^{2}}\;,  \label{eq-k28}
\end{equation}%
and an important constraint in the reaction is the following:%
\begin{equation}
|y|\leq p\leq Y\;.  \label{eq-k29}
\end{equation}%
Furthermore the cosine of the angle $\theta _{pq}$ in \cite{semi} is given by%
\begin{equation}
\cos \theta _{pq}=\frac{1}{2pq}\left[ W^{2}-2(M_{d}+\omega )\sqrt{%
m_{p}^{2}+p^{2}}\right] \;.  \label{eq-k30}
\end{equation}%

\section{Weak responses and cross section \label{formalism_crosssection}}

Using the Feynman rules and integrating over the undetected nucleon three-momentum, the cross section of the process can be written as
\begin{eqnarray}
d\sigma_{\chi} = \frac{G^2\:\cos^2 \theta_c}{2(2\pi)^5}
\:\frac{m_p^2 \:v_0}{k \:\varepsilon^{\prime} \:E_1 \:E_2}
\:\mathcal{F}^2_{\chi} \:d^3\mathbf{k^{\prime}} \:d^3\mathbf{p_1} \:\delta(\varepsilon+M_d-\varepsilon^{\prime}-E_1-E_2) \;, \label{cross_section_general}
\end{eqnarray}
where  $\mathcal{F}^2_{\chi}$ is the matrix element squared (see below), $v_0 \equiv (\varepsilon +\varepsilon^{\prime})^2-q^2$ and the differentials can be expressed as $d^3\mathbf{k^{\prime}} = k^{\prime 2} \:dk^{\prime} \:d\Omega_{k^{\prime}} = k^{\prime} \:\varepsilon^{\prime} \:d\varepsilon^{\prime} \:d\Omega_{k^{\prime}}$ and $d^3\mathbf{p_1} = p_1^2 \:dp_1 \:d\Omega_{p_1} = p_1 \:E_1 \:dE_1 \:d\Omega_{p_1}$. From this expression one can construct the differential cross section with respect to any set of variables, taking into account that the integration of the delta function $\delta(f(x,...))$ with respect to the variable $x$ forces the variables of $f$ to fulfill the condition $f\equiv 0$ and introduces an extra factor $|\partial f(x,...) / \partial x|^{-1}$.

As an example, if the energy distribution of the incoming neutrino beam, $P(\varepsilon)$, is known, the differential cross section averaged over incident energies is given by
\begin{equation}\label{cross_section_energydistribution}
\int{\frac{d\sigma_{\chi}}{d\varepsilon^{\prime }\:d\Omega _{k^{\prime }}\:dE_1\:d\Omega _{p_1}}} \:P(\varepsilon) \:d\varepsilon =
\frac{G^2\:\cos^2 \theta_c}{2(2\pi)^5}
\:m_p^2 \:\frac{p_1 \:k^{\prime} \:v_0}{k \:E_2} 
\:P(\varepsilon) \:|F_P|^{-1}  \:\mathcal{F}^2_{\chi}\;,
\end{equation}
with
\begin{equation}
F_P = 1 - \frac{\varepsilon \:\left( 1-\frac{p_1\:\cos\theta_1}{q} \right) \:\left( 1-\frac{k^{\prime}\:\cos\theta}{k} \right)}{E_2} \;.
\end{equation}

As another example, by a further integration of Eq.~(\ref{cross_section_general}) over the undetected nucleon energy one gets the following differential cross section:
\begin{equation}\label{cross_section}
\frac{d\sigma_{\chi}}{dk^{\prime }\:d\Omega _{k^{\prime }}\:d\Omega _{p_1}} =
\frac{G^2\:\cos^2 \theta_c}{2(2\pi)^5}
\:\frac{m_p^2}{M_d}
\:\frac{p_1 \:k^{\prime 2} \:v_0}{k \:\varepsilon^{\prime}} \:|F|^{-1} \:\mathcal{F}^2_{\chi}
\end{equation}
with
\begin{equation}
F = 1 + \frac{\omega\:p_1 - q\:E_1\:\cos\theta_1}{M_d\:p_1}\;,
\end{equation}
which becomes in extreme relativistic limit
\begin{equation}\label{cross_section_erl}
\frac{d\sigma_{\chi \:\text{[ERL]}}}{d\varepsilon^{\prime }\:d\Omega _{k^{\prime }}\:d\Omega _{p_1}} =
\frac{G^2\:\cos^2 \theta_c}{16\pi^5}
\:\frac{m_p^2}{M_d}
\:p_1 \:\varepsilon^{\prime 2} \:\cos^2(\theta/2)  \:|F|^{-1} \:\mathcal{F}^2_{\chi}\;.
\end{equation}
The energy and momenta in these expressions fulfill conservation laws and thus the undetected energies $\varepsilon$ and $E_2$ (or momenta $k$ and $p_2$) take only specific values that can be deduced as described in the previous section.

The matrix element squared is given by the contraction of the leptonic and hadronic tensors, which can be written in terms of products of generalized Rosenbluth factors $V$ and hadronic responses $w$ for charge, longitudinal and transverse projections:
\begin{eqnarray}\nonumber
\mathcal{F}^2_{\chi} &=& {V}_{CC} \:(w^{VV}_{CC}+w^{AA}_{CC}) + {V}_{CL} \:(w^{VV}_{CL}+w^{AA}_{CL}) + {V}_{LL} \:(w^{VV}_{LL}+w^{AA}_{LL}) + \\\nonumber
&& {V}_{T} \:(w^{VV}_{T}+w^{AA}_{T}) + {V}_{TT} \:(w^{VV}_{TT}+w^{AA}_{TT}) + {V}_{TC} \:(w^{VV}_{TC}+w^{AA}_{TC}) + \\
&& + {V}_{TL} \:(w^{VV}_{TL}+w^{AA}_{TL}) + \chi \left[ {V}_{T^{\prime }}w^{VA}_{T^{\prime }}+{V}_{TC^{\prime }}w^{VA}_{TC^{\prime }}+{V}_{TL^{\prime }}w^{VA}_{TL^{\prime }} \right] \;,
\label{matrix_element}
\end{eqnarray}
where $\chi=+1$ for neutrino scattering and $\chi=-1$ for antineutrino scattering. The generalized Rosenbluth factors contributing to this matrix element are given in \cite{semi} for general lepton masses using the quantities $\delta  \equiv m/\sqrt{|Q^{2}|}$ and $\delta^{\prime }\equiv m^{\prime}/\sqrt{|Q^{2}|}$ and for general values of the vector and axial-vector coupling constants $a_V$ and $a_A$.  Here we reproduce those expressions particularized to the Standard Model tree-level values of the coupling constants ($a_V=-a_A=1$) and to (anti)neutrino scattering with $m=0$:
\begin{eqnarray}
{V}_{CC} & = & 1 - \delta^{\prime 2} \:\tan^{2}\widetilde{\theta }/2 \label{vcc} \\
{V}_{CL} & = & - \nu - \frac{1}{\rho ^{\prime }} \:\delta ^{\prime 2} \:\tan ^{2}\widetilde{\theta }/2 \label{vcl} \\
{V}_{LL} & = &  \nu ^{2}+ \left(1+ \frac{2\nu}{\rho ^{\prime }} + \rho \:\delta ^{\prime 2}\right) \:\delta ^{\prime 2} \:\tan ^{2}\widetilde{\theta }/2  \label{vll} \\
{V}_{T} & = & \frac{1}{2}\rho + \left( 1-\frac{\nu }{\rho ^{\prime }} \:\delta^{\prime 2} -\frac{1}{2}\rho \:\delta ^{\prime 4} \right) \tan ^{2}\widetilde{\theta }/2  \label{vt}\\
{V}_{TT} & = & -\frac{1}{2}\rho +\left[ 1 +\frac{\nu }{\rho ^{\prime }} +\frac{1}{2}\rho \:\delta ^{\prime 2}\right] \delta ^{\prime 2} \:\tan ^{2}\widetilde{\theta }/2\label{vtt} \\
{V}_{TC} & = & - \frac{1}{\rho ^{\prime }}\tan \widetilde{\theta }/2 \:\sqrt{-\frac{1}{\rho}\:{V}_{TT}} \label{vtc} \\
{V}_{TL} & = & -\left( \nu +\rho \rho ^{\prime }\:\delta^{\prime 2} \right) \:{V}_{TC}  \label{vtl} \\
{V}_{T^{\prime }} & = &  \left( -\frac{1}{\rho^{\prime }} + \nu\:\delta^{\prime 2} \right) \:\tan ^{2}\widetilde{\theta }/2 \label{vtp} \\
{V}_{TC^{\prime }} & = & \tan \widetilde{\theta }/2 \left\{ \frac{1}{2}-\frac{1}{\rho }\left[ 1 +\frac{\nu }{\rho ^{\prime }} +\frac{1}{2}\rho \:\delta ^{\prime 2}\right] \delta^{\prime 2} \:\tan ^{2}\widetilde{\theta }/2\right\}^{1/2}   \label{vtcp} \\
{V}_{TL^{\prime }} & = & -\nu {V}_{TC^{\prime }} \;. \label{vtlp}
\end{eqnarray}
The following definitions have been used in the expressions above:
\begin{equation}
\begin{array}{llll}
\nu \equiv \frac{\omega }{q} \qquad & \tan^{2}\widetilde{\theta }/2 \equiv \frac{|Q^2|}{v_0} \qquad & \rho  \equiv \frac{|Q^{2}|}{q^{2}} \qquad & \rho^{\prime } \equiv \frac{q}{\varepsilon +\varepsilon^{\prime}} \;.
\end{array}
\label{definitions}
\end{equation}

The deuteron responses $w$, on the other hand, can be constructed as functions of the three independent four-momenta of the hadronic part of the process, namely $Q^{\mu}$, $P_d^{\mu}$, and $P_1^{\mu}$, in 17 different $VV$, $AA$ and $VA$ combinations, each associated to an invariant function of the four dynamical invariants constructed with the same set of four-momenta \cite{semi}. The computation of these deuteron responses will be described in the next section for a sophisticated model of the nucleon structure.

\section{Deuteron structure \label{formalism_deuteron}}

The calculation of matrix elements for electro-weak breakup of the deuteron requires that some care be exercised in the construction of a consistent model of the reaction in order that basic symmetries, such as electromagnetic current conservation, be maintained. This is a problem that has been discussed by many authors in the context of deuteron electrodisintegration. A detailed discussion in the context of the Bethe-Salpeter equation \cite{Salpeter:1951sz} and the related spectator equation \cite{Gross:1982nz,Gross:1982ny} can be found in \cite{Adam:1997cx,Adam:2004yt}. In general, for all approaches a nucleon-nucleon ($NN$) interaction is parameterized in terms of a potential or interaction kernel which can be iterated in the appropriate equation, such as the Schr\"odinger equation or Bethe-Salpeter equation, to produce the deuteron bound-state wave function and scattering amplitudes which is then fit to data for laboratory kinetic energies up to $T_{lab}\sim$ 350 MeV, which is slightly above pion production threshold. Since the $NN$ interaction generally involves the exchange of electro-weak charges, it is necessary in constructing the model to include two-body currents that contain coupling to the exchanged particles that carry the charges. This program generally relies on the ability to relate the $NN$ interaction to some set of effective meson exchanges and has been carried out rigorously, for example, in the context of chiral effective field theory \cite{Girlanda:2014yea} and the covariant spectator equation \cite{Gross:2007jj,Gross:2014wqa,Gross:2014tya}. However, attempts to extend this approach to invariant masses well above pion threshold have been relatively unsuccessful and at large four-momentum transfers and invariant masses it is necessary to construct models which are not fully consistent as a result of the inability to produce interaction kernels that can successfully describe $NN$ scattering in this region. 

The calculations presented in this paper use a model related to the Bethe-Salpeter equation, which is designed to be used at large four-momentum transfers and invariant masses and has been used previously to describe deuteron electrodisintegration \cite{JVO_2008_newcalc,JVO_2009_tar_pol,JVO_2009_ejec_pol,FJVO,Ford:2014yua}. A brief outline of the model containing the modifications required for calculation of the CC$\nu$ reaction will be presented here. This model is constructed primarily for use at large $Q^2$ where relativistic effects are important and where there are open meson-production channels that must appear as inelasticities in the scattering matrix producing the final-state interaction. However, the plane-wave contributions can be used at all values of $Q^2$. At this stage no two-body meson-exchange currents (MEC) are included in the calculations, although these will be included in the future, meaning that the calculations shown here are in the impulse approximation (IA). As a rough estimate of what this approximation might entail, we note that in \cite{VanOrden:1980tg} the scaling behavior of EM (vector) MEC was found to be such that the results of the ratio between MEC and IA goes as $k_A^4$, where $k_A$ is some characteristic three-momentum for a given nucleus (roughly the Fermi momentum $k_F$). This ratio is typically found to be about 15-20\% at the maxima of the quasielastic and MEC contributions for nuclei such as $^{12}$C where $k_A \cong$ 228 MeV. However, the characteristic momentum for $^2$H is about 55 MeV, which yields a rough estimate for this ratio in the present situation of $\sim$5--7$\times 10^{-4}$, namely, contributions from MEC should be very small. As a result, this calculation is limited to the IA. 

\begin{figure}
\centerline{\includegraphics[height=2.5in]{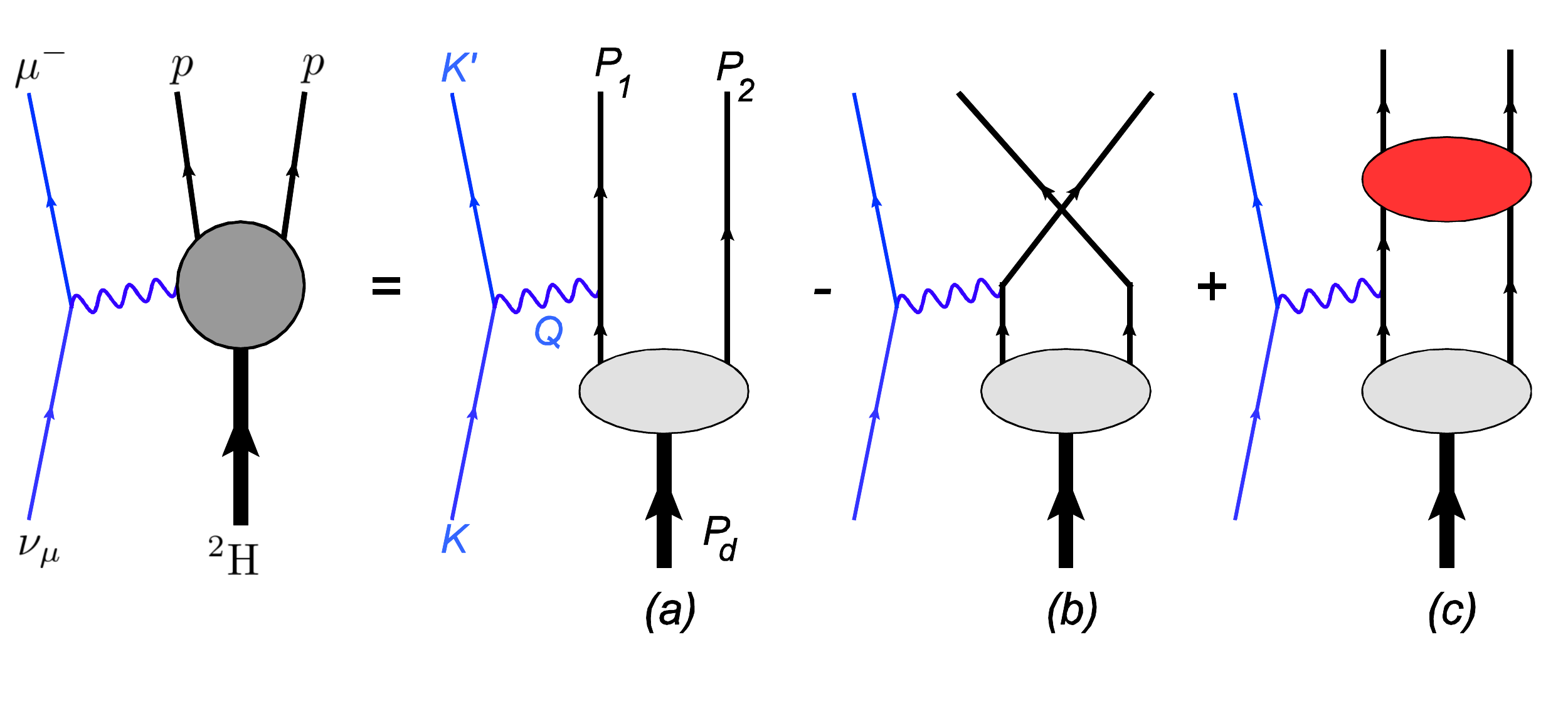}}
\caption{(Color online) Feynman diagrams representing charge-changing neutrino reactions in the impulse
approximation.  }\label{fig:impulse}
\end{figure}

The Feynman diagrams representing the impulse approximation are
shown in Fig. \ref{fig:impulse}.
Fig. \ref{fig:impulse}a represents the direct plane-wave contribution, 
Fig. \ref{fig:impulse}b represents the corresponding exchange contribution and 
Fig. \ref{fig:impulse}c represents the contribution from final-state
interactions. Use of Fig. \ref{fig:impulse}a alone is called the plane-wave impulse approximation (PWIA), while adding the exchange diagram Fig. \ref{fig:impulse}b gives the plane-wave Born approximation (PWBA) \cite{Arenhovel:1982rx,Poulis:1993wm}. We refer to the inclusion of all three diagrams in Fig. \ref{fig:impulse} as the distorted-wave Born approximation (DWBA). In the next section results are shown for all three assumptions to ascertain the impact of antisymmetrization and final-state interactions on the predictions made in this study.

The matrix element corresponding to \ref{fig:impulse}a is given by
\begin{equation}
\left<\mathbf{p}_1s_1;\mathbf{p}_2s_2\right|J^\mu(Q)\left|\mathbf{p_d}\lambda_d\right>_{a}
=-\bar{u}(\mathbf{p}_1,s_1)\Gamma^\mu_{CC}(q)G_0(P_d-P_2)
\Gamma^T_{\lambda_d}(P_2,P_d)\bar{u}^T(\mathbf{p}_2,s_2)\,,
\end{equation}
where the target deuteron has four-momentum $P_d$ and spin
$\lambda_d$, the two final-state protons have four-momentum and spin $(P_1,s_1)$
and  $(P_2,s_2)$. The
single-nucleon propagator is
\begin{equation}
G_0(P)=\frac{\gamma\cdot P+m_N}{m_N^2-P^2-i\eta}
\end{equation}
and the weak charged-current operator has vector and axial-vector contributions
\begin{equation}
\Gamma_{CC}^\mu(Q)=[\Gamma_V^\mu(Q)-\Gamma_A^\mu(Q)]\tau_+
\end{equation}
\begin{equation}
\Gamma_V^\mu(Q)=F_1^{I=1}(Q^2)\gamma^\mu+\frac{F_2^{I=1}(Q^2)}{2m_N}i\sigma^{\mu\nu}Q_\nu
\end{equation}
\begin{equation}
\Gamma_A^\mu(Q)=\left[G_A(Q^2)\gamma^\mu+G_P(Q^2)\frac{Q^\mu}{2m_N}\right]\gamma_5\,,
\end{equation}
where the vector-isovector form factors used here are from the GKex05 parameterization (see \cite{Crawford:2010gv} and references therein) of the nucleon electromagnetic form factors and the axial-vector and induced pseudoscalar form factors are taken to be
\begin{equation}
G_A(Q^2)=\frac{1.2695}{\left( 1+\frac{Q^2}{M_A^2}\right)^2} \label{axial_ff}
\end{equation}
and
\begin{equation}
G_P(Q^2)=\frac{1}{\left( \frac{1}{185.05}+\frac{Q^2}
{4m_p^2}\right)}G_A(Q^2)\,.  \label{pseudoscalar_ff}
\end{equation}
Unless stated otherwise we use $M_A=1.03\ {\rm GeV}$.

For this contribution, the one leg of the deuteron is on-shell.
The deuteron vertex function with nucleon 2 on-shell can be written
as
\begin{eqnarray}
&& \Gamma_{\lambda_d}(P_2,P_d) = g_1(P_2^2,P_2\cdot
P_d)\gamma\cdot\xi_{\lambda_d}(P) +g_2(P_2^2,P_2\cdot
P_d)
\frac{P\cdot\xi_{\lambda_d}(P_d)}{m_N}\nonumber\\
&&-\left(g_3(P_2^2,P_2\cdot
P_d)\gamma\cdot\xi_{\lambda_d}(P_d)
+g_4(P_2^2,P_2\cdot
P_d)\frac{P\cdot\xi_{\lambda_d}(P_d)}{m_N}\right)\frac{\gamma\cdot
p_1+m}{m_N}C\,,
\end{eqnarray}
where $P_1=P_d-P_2$, $P=\frac{1}{2}(P_1-P_2)=\frac{P_d}{2}-P_2$, $C$ is
the charge-conjugation matrix and $\xi_{\lambda_d}$ is the deuteron
polarization four-vector. The invariant functions $g_i$ are given by

\begin{eqnarray}
g_1(P_2^2,P_2\cdot P_d)&=&\frac{2E_{\kappa}-M_d}{\sqrt{8\pi}}\left[ u(\kappa)-\frac{1}{\sqrt{2}}w(\kappa)+\sqrt{\frac{3}{2}}\frac{m_N}{\kappa}v_t(\kappa)\right]\\
g_2(P_2^2,P_2\cdot P_d)&=&\frac{2E_{\kappa}-M_d}{\sqrt{8\pi}}\left[ \frac{m_N}{E_{\kappa}+m_N}u(\kappa)+\frac{m_N(2E_{\kappa}+m_N)}{\sqrt{2}\kappa^2}w(\kappa)+\sqrt{\frac{3}{2}}\frac{m_N}{\kappa}v_t(\kappa)\right]\\
g_3(P_2^2,P_2\cdot P_d)&=&\sqrt{\frac{3}{16\pi}}\frac{m_N E_{\kappa}}{\kappa}v_t(\kappa)\\
g_4(P_2^2,P_2\cdot P_d)&=&-\frac{m_N^2}{\sqrt{8\pi}M_d}\left[(2 E_{\kappa}-M_d)\left(\frac{1}{E_{\kappa}+m_N}u(\kappa) -\frac{E_{\kappa}+2m_N}{\sqrt{2}\kappa^2}w(\kappa)\right)+\frac{\sqrt{3}M_d}{\kappa}v_s(\kappa)\right]\,,\nonumber\\
\end{eqnarray}
where
\begin{equation}
\kappa=\sqrt{\frac{(P_d\cdot P_2)^2}{P_d^2}-P_2^2}
\end{equation}
is the magnitude of the neutron three-momentum in the deuteron rest
frame and
\begin{equation}
E_{\kappa}=\sqrt{\kappa^2+m_N^2}\,.
\end{equation}
The functions $u(\kappa)$, $w(\kappa)$, $v_s(\kappa)$ and $v_t(\kappa)$ are the s-wave,
d-wave, singlet p-wave and triple p-wave radial wave functions of
the deuteron in momentum space.  
For convenience, the spectator deuteron wave function can be defined as
\begin{equation}
\psi_{\lambda_d,s_2}(P_2,P_d)=G_0(P_d-P_2)
\Gamma^T_{\lambda_d}(P_2,P_d)\bar{u}^T(\mathbf{p}_2,s_2)\,.
\end{equation}
We choose to normalize this wave function such that in the deuteron
rest frame
\begin{equation}
\sum_{s_2}\int\frac{d^3p_2}{(2\pi)^3}\frac{m_N}{E_{p_2}}\bar{\psi}_{\lambda_d,s_2}(P_2,P_d)
\gamma^0 \psi_{\lambda_d,s_2}(P_2,P_d)=1\,,
\end{equation}
which is correct only in the absence of energy-dependent kernels \cite{Adam:1997cx}.
This corresponds to a normalization of the radial wave functions 
\begin{equation}
\int_0^\infty
\frac{d\kappa\kappa^2}{(2\pi)^3}\left[u^2(\kappa)+w^2(\kappa)+v_t^2(\kappa)+v_s^2(\kappa)\right]=1\,.
\end{equation}
For the calculations in this paper we use the WJC 2 wave functions \cite{Gross:2007jj}.

The current matrix element corresponding to Fig. \ref{fig:impulse}a can then
be written as
\begin{equation}
\left<\mathbf{p}_1s_1;\mathbf{p}_2s_2\right|J^\mu(Q)\left|\mathbf{p}_d\lambda_d\right>_a=
-\bar{u}(\mathbf{p}_1,s_1)\Gamma^\mu_{CC}(Q)\psi_{\lambda_d,s_2}(P_2,P_d) \;.
\end{equation}

The current matrix element corresponding to Fig. \ref{fig:impulse}b is related to that of Fig. \ref{fig:impulse}a by a crossing of the two final-state protons. So,
\begin{equation}
\left<\mathbf{p}_1s_1;\mathbf{p}_2s_2\right|J^\mu(Q)\left|\mathbf{p_d}\lambda_d\right>_b
=\left<\mathbf{p}_2s_2;\mathbf{p}_1s_1\right|J^\mu(Q)\left|\mathbf{p_d}\lambda_d\right>_a\,.
\end{equation}

The contribution from final-state interactions represented by Fig. \ref{fig:impulse}c requires the introduction of a $pp$ scattering amplitude $M$ and an integration for the loop four-momentum
$k_2$, which involves both the deuteron vertex function and the $pp$
scattering amplitude. In this case, both of the protons are in general off-shell.  However, an examination of the contributions of the poles of the nucleon propagators shows that the contribution from the positive-energy on-shell pole of proton 2 dominates the calculation. As in the previous electrodisintegration calculations, we choose to put particle 2 on its positive-energy mass shell to simplify the calculation.  Using this approach, the contribution of the final-state
interaction to the current matrix element is given by
\begin{eqnarray}
\left<\mathbf{p}_1s_1;\mathbf{p}_2s_2\right|J^\mu(Q)\left|\mathbf{p}_d\lambda_d\right>_c&=&\int
\frac{d^3k_2}{(2\pi)^3}\frac{m}{E_{k_2}}
\bar{u}_a(\mathbf{p}_1,s_1)\bar{u}_b(\mathbf{p}_2,s_2)M_{ab;cd}(P_1,P_2;K_2)
\nonumber\\
&&\times{G_0}_{ce}(P_d+Q-K_2)\Gamma^\mu_{CCef}(q){G_0}_{fg}(P_d-K_2)\nonumber\\
&&\times\Lambda^+_{dh}(\mathbf{k}_2){\Gamma^T_{\lambda_d}}_{gh}(K_2,P)\,,\label{J_FSI}
\end{eqnarray}
where $M$ is the $pp$ scattering amplitude,
\begin{equation}
\Lambda^+(\mathbf{p})=\sum_s
u(\mathbf{p},s)\bar{u}(\mathbf{p},s)=\frac{\gamma\cdot P+m}{2m}
\end{equation}
is the positive energy projection operator and the Dirac indices for
the various components are shown explicitly.

At this point, only the incoming momentum for particle 1 is off-shell. For small values of the final-state invariant mass, this poses no problem, since the spectator equation could be used to construct a scattering matrix consistent with the deuteron bound state and a consistent current operator could also be constructed. However, for invariant mass well above pion threshold, there are no existing particle-exchange models of the kernel that reproduce the data. For this reason, it is necessary to use scattering matrices that have been fit to data, which limits the scattering amplitudes to the case where all legs  of the scattering matrix are on-mass-shell. In \cite{JVO_2008_newcalc}, the separation of the propagator for particle 1 into on-shell and off-shell contributions is described in some detail.  All contributions can be calculated if a prescription is provided for taking the initial four-momentum of particle 1 off-mass-shell. This problem has been discussed in \cite{Ford:2013zca}. In the calculations presented here, we will use only the on-shell contribution which is well determined and is dominant.  

The completely on-mass-shell scattering amplitude can be parameterized in terms of five Fermi
invariants as
\begin{eqnarray}
M_{ab;cd}&=&\mathcal{F}_S(s,t)\delta_{ac}\delta_{bd}+\mathcal{F}_V(s,t)\gamma_{ac}
\cdot\gamma_{bd}+\mathcal{F}_T(s,t)\sigma^{\mu\nu}_{ac}(\sigma_{\mu\nu}) _{bd}^{} \nonumber\\
&&+\mathcal{F}_{P}(s,t)\gamma^5_{ac}\gamma^5_{bd}+
\mathcal{F}_A(s,t)(\gamma^5\gamma)_{ac}\cdot(\gamma^5\gamma)_{bd} \;, \label{Fermi}
\label{eqdefnn}
\end{eqnarray}
where $s$ and $t$ are the usual Mandelstam variables and amplitudes $\mathcal{F}_i$ can be separated into isoscalar and isovector contributions using
\begin{equation}
\mathcal{F}_i(s,t)=\mathcal{F}_i^{I=0}(s,t)+\mathcal{F}_i^{I=1}(s,t)\boldsymbol{\tau}^{(1)}\cdot\boldsymbol{\tau}^{(2)}\,.
\end{equation}
The invariant amplitudes have been constructed in two ways. First, the helicity amplitudes obtained from SAID \cite{Arndt00,Arndt07,SAIDdata} have been used to obtain the invariant functions as described in \cite{JVO_2008_newcalc}. In this case, the analysis of the $pn$ amplitudes is limited to $s\leq 5.97\ {\rm GeV}^2$ and the $pp$ are limited to $s\leq 9.16\ {\rm GeV}^2$. The second method is the use of a Regge model of $NN$ scattering fit to cross sections and spin observables over the range $5.4\ {\rm GeV}^2\leq s \leq 4000\ {\rm GeV}^2$ \cite{FVO_Reggemodel,Ford:2013wxa}.

With this brief summary of the dynamics used in the present study, let us now proceed to the presentation of some typical results both for inclusive and semi-inclusive CC$\nu$ reactions with deuteron.

\section{Results \label{results}}

\subsection{Inclusive cross sections}

\begin{figure}
\begin{center}
\includegraphics[width=0.95\textwidth]{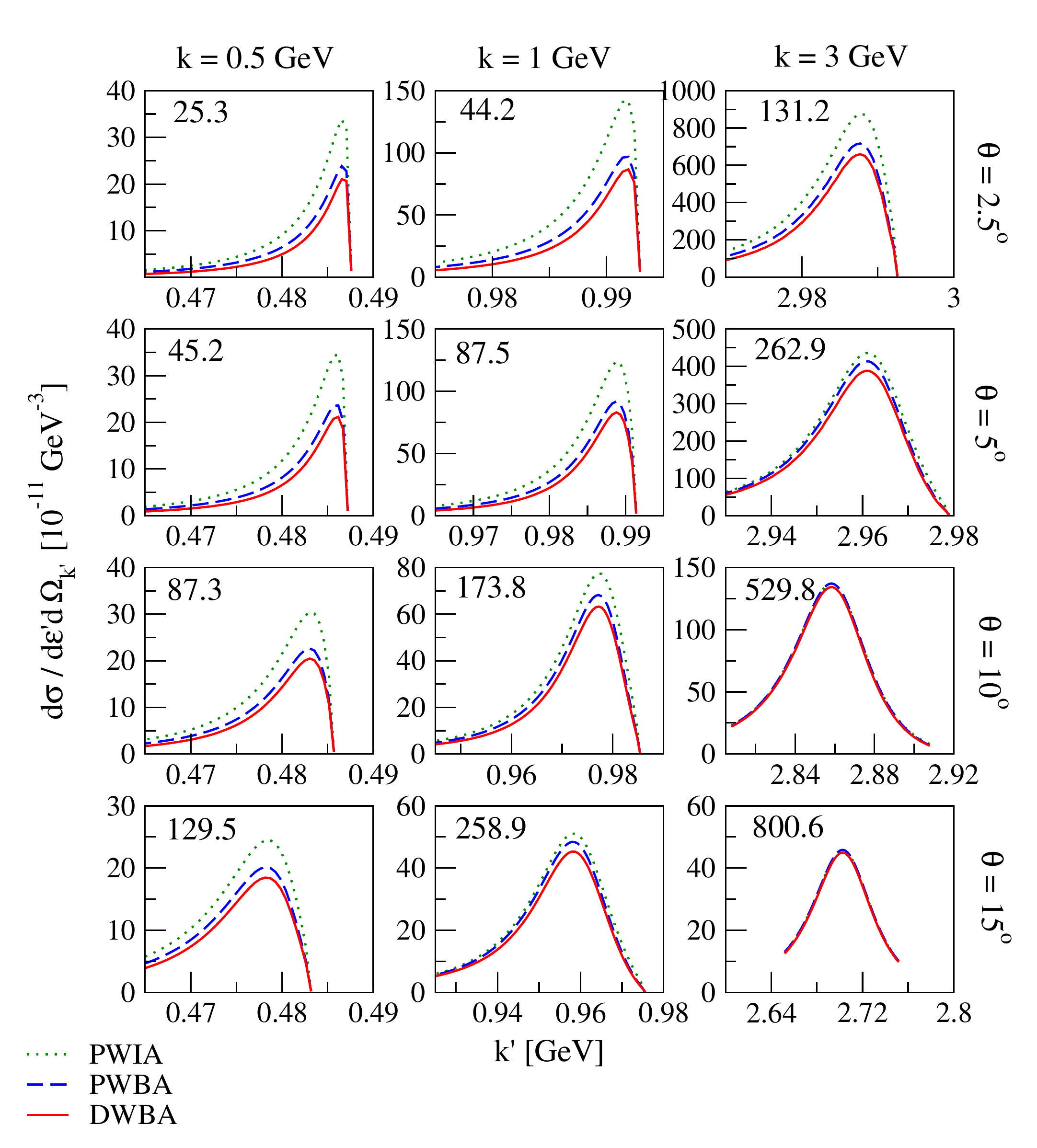}
\caption{(Color online) CC$\nu$ deuteron disintegration inclusive no-pion cross section as a function of the outgoing muon momentum $k'$ for three incident neutrino momenta ($k=$ 0.5, 1 and 3 GeV) and four lepton scattering angles ($\theta=$ 2.5$^{\text{o}}$, 5$^{\text{o}}$, 10$^{\text{o}}$, and 15$^{\text{o}}$). Dotted lines correspond to PWIA results, dashed lines to PWBA and solid lines to DWBA. Numbers in the top left corner of each plot correspond to the momentum transfer $q$ in MeV at the peak of the cross sections.
\label{inclusive_deuteron1}}
\end{center}
\end{figure}

\begin{figure}
\begin{center}
\includegraphics[width=0.95\textwidth]{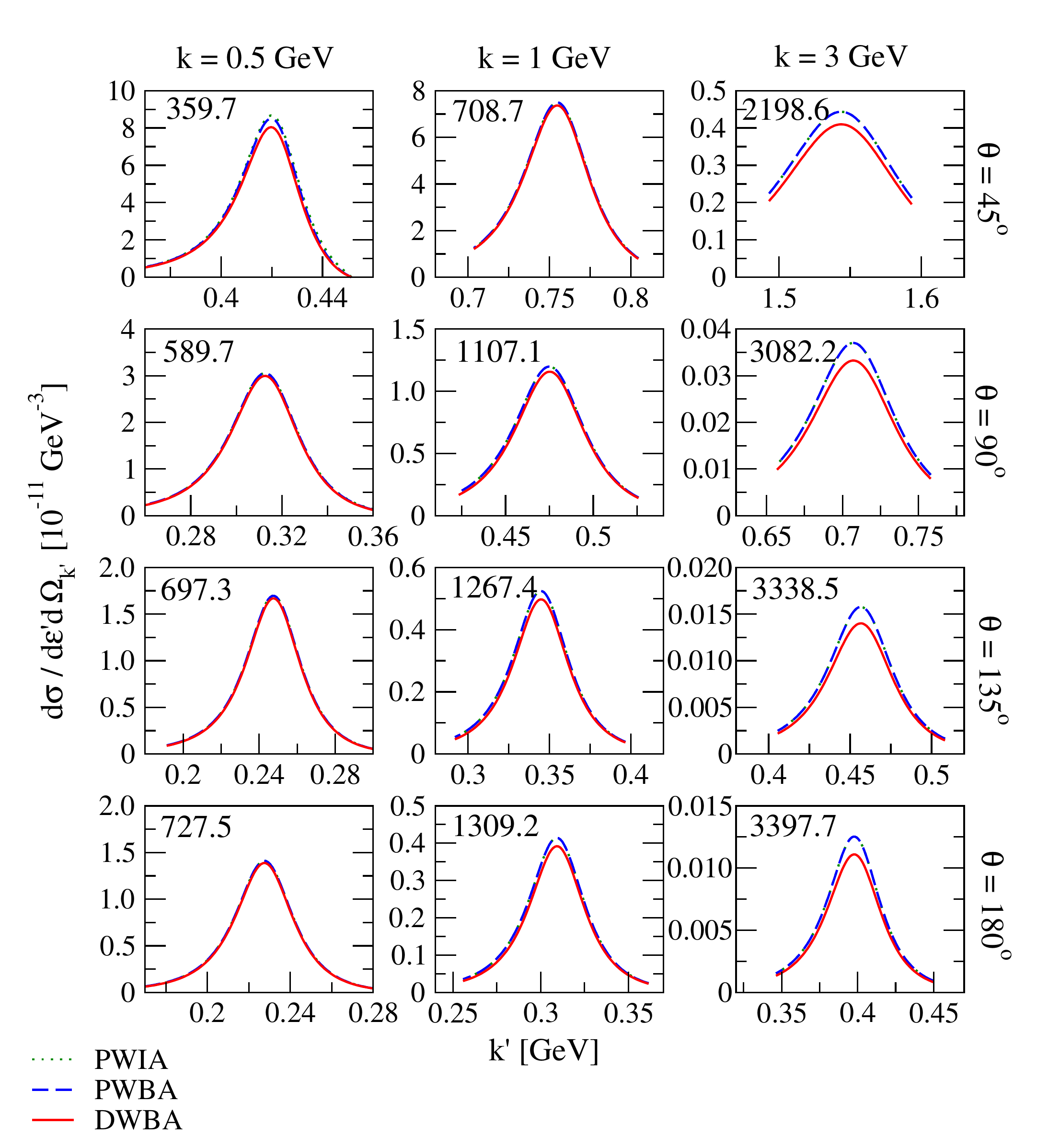}
\caption{(Color online) As for Fig. \ref{inclusive_deuteron1}, but now for larger scattering angles ($\theta=$ 45$^{\text{o}}$, 90$^{\text{o}}$, 135$^{\text{o}}$, and 180$^{\text{o}}$).
\label{inclusive_deuteron2}}
\end{center}
\end{figure}

We start by considering results for the inclusive cross section, {\it i.e.,} where only the final-state charged lepton is presumed to be detected. We assume that only the no-pion cross section is allowed and that other channels are not involved. This, of course, is an issue for experimental studies and any inclusive measurement must involve all open channels. Our purpose in starting with modeling of the inclusive no-pion cross section is to ascertain where in $q$, $\omega$ and scattering angle one should expect to find significant CC$\nu$ strength before going on to study the semi-inclusive reaction. We begin by showing in Figs. \ref{inclusive_deuteron1} and \ref{inclusive_deuteron2} the inclusive no-pion CC$\nu$ deuteron disintegration cross sections as a function of the momentum of the outgoing muon for three incident neutrino momenta ($k=$ 0.5, 1 and 3 GeV) and for four small lepton scattering angles, $\theta=$ 2.5$^{\text{o}}$, 5$^{\text{o}}$, 10$^{\text{o}}$, and 15$^{\text{o}}$ (Fig. \ref{inclusive_deuteron1}) and four larger angles, $\theta=$ 45$^{\text{o}}$, 90$^{\text{o}}$, 135$^{\text{o}}$, and 180$^{\text{o}}$ (Fig. \ref{inclusive_deuteron2}). In each plot the corresponding momentum transfer at the peak (in MeV) is indicated, and results for PWIA, PWBA and DWIA calculations are shown. From these curves we extract the muon momentum at the peak of the inclusive no-pion cross section ($k'_0$), as well as at one order of magnitude below the peak before and after $k'_0$ ($k'_{<}$ and $k'_{>}$, respectively).
As can be seen in these plots, the general trend is a decrease of the cross section when including exchange terms, {\it i.e.}, from PWIA to PWBA, as well as when considering final-state interactions, {\it i.e.}, from PWBA to DWBA. For momentum transfers around 500 MeV and larger the PWIA and PWBA results are indistinguishable; and they are also indistinguishable from the DWBA result for momentum transfers between 500 and 1000 MeV, approximately. From the last we conclude that, at least for inclusive no-pion scattering when at the peak of the cross section and when the momentum transfer is reasonably large, the details of the modeling ({\it i.e.,} as reflected in the three models used in the present work) are not too important. In contrast, for kinematics where the exchanged momentum is small, there are significant differences in going from model to model, and for quantitative comparisons with data some care should be exercised.

\subsection{Semi-inclusive cross sections}

\begin{figure}
\begin{center}
\includegraphics[width=0.98\textwidth]{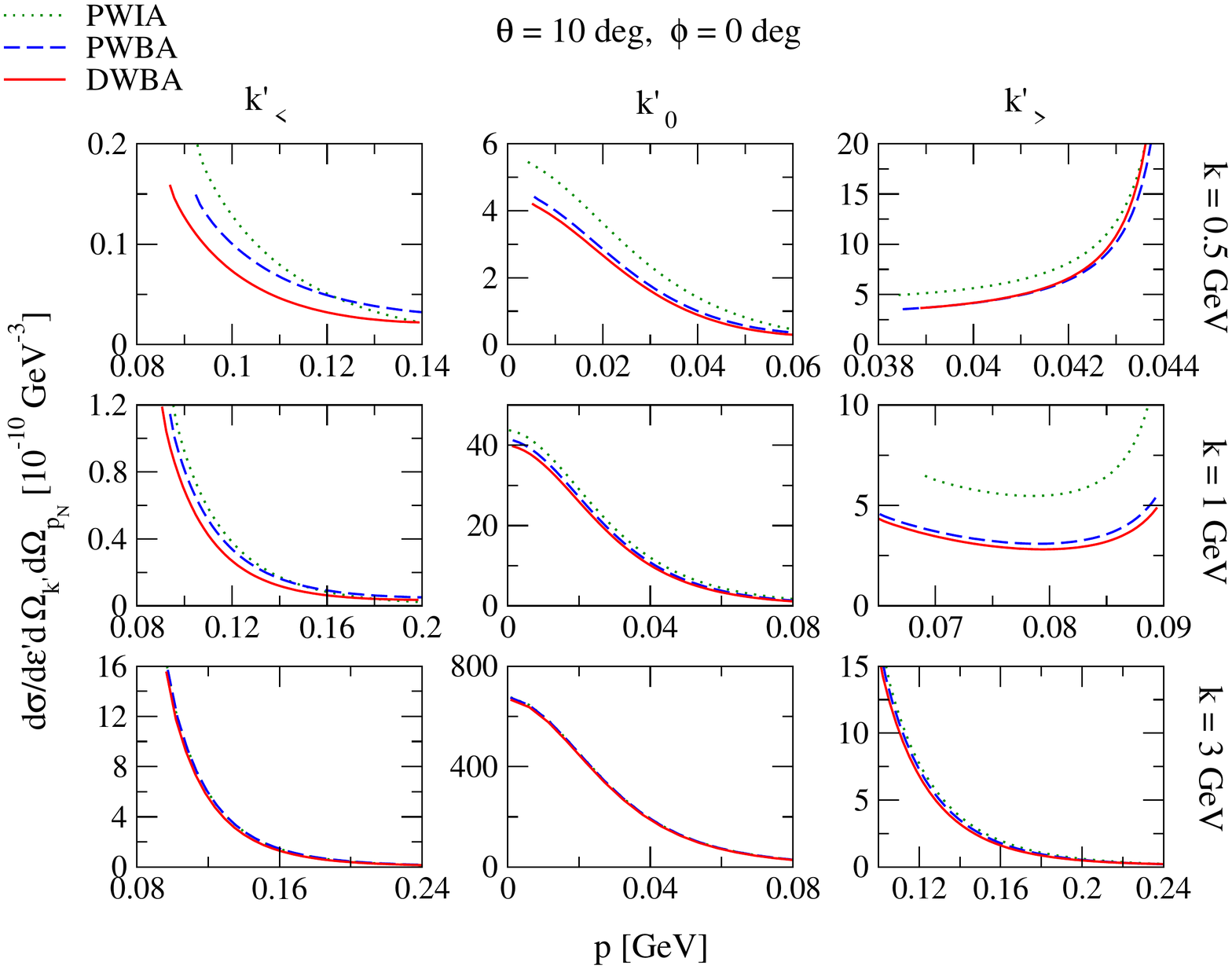}
\caption{(Color online) CC$\nu$ deuteron disintegration semi-inclusive cross section as a function of the missing momentum $p$ for three incident neutrino momenta ($k=$ 0.5, 1, and 3 GeV), and three outgoing muon momenta ($k'=k'_{<}$, $k'_0$, and $k'_{>}$ , see text). The lepton scattering angle is fixed to $\theta=$ 10$^{\text{o}}$ and nucleon emission angle to $\phi=$ 0$^{\text{o}}$. Results correspond to PWIA (dotted line), PWBA (dashed line) and DWBA (solid line).
\label{semiinclusive_phi0_th10}}
\end{center}
\end{figure}

\begin{figure}
\begin{center}
\includegraphics[width=0.98\textwidth]{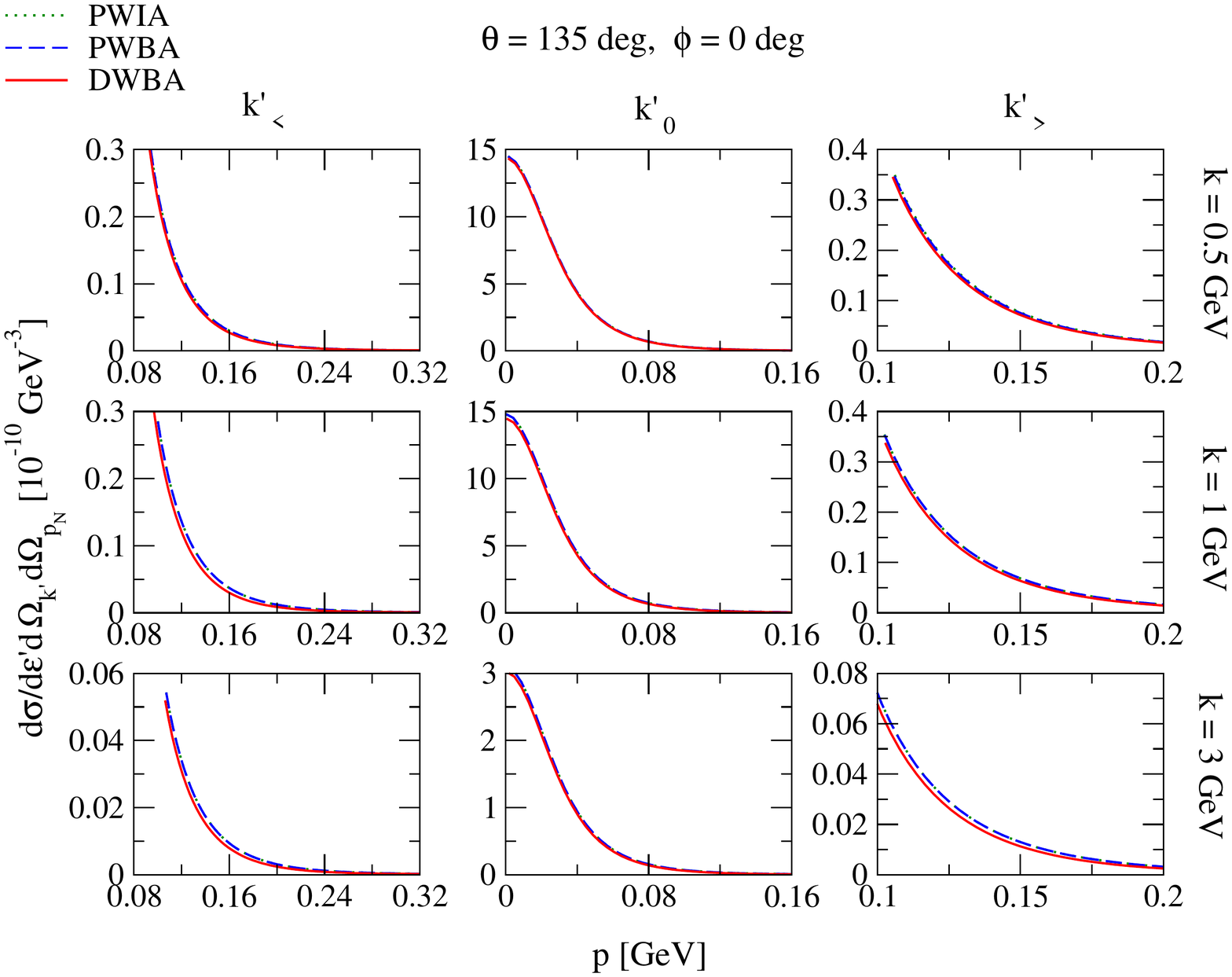}
\caption{(Color online) As for Fig. \ref{semiinclusive_phi0_th10}, but now for lepton scattering angle $\theta=$ 135$^{\text{o}}$.
\label{semiinclusive_phi0_th135}}
\end{center}
\end{figure}

For each combination of initial and final lepton momenta, $k=$ 0.5, 1 and 3 GeV and $k'=k'_{<}, k'_0, k'_{>}$ (the latter dependent on $k$ and $\theta$), we show in Figs. \ref{semiinclusive_phi0_th10} and  \ref{semiinclusive_phi0_th135} CC$\nu$ deuteron disintegration semi-inclusive cross sections as a function of the missing momentum (residual nucleon momentum) for two representative lepton scattering angles, $\theta=$ 10$^{\text{o}}$ and 135$^{\text{o}}$, and one representative nucleon emission angle $\phi=$ 0$^{\text{o}}$. We show PWIA, PWBA and DWBA results, which for $\theta=$ 0$^{\text{o}}$ follow the same trend as for the inclusive no-pion cross sections, namely decreasing values in the order PWIA, PWBA, DWBA. For $\theta=$ 135$^{\text{o}}$ the three calculations are very close.

\begin{figure}
\begin{center}
\includegraphics[width=0.8\textwidth]{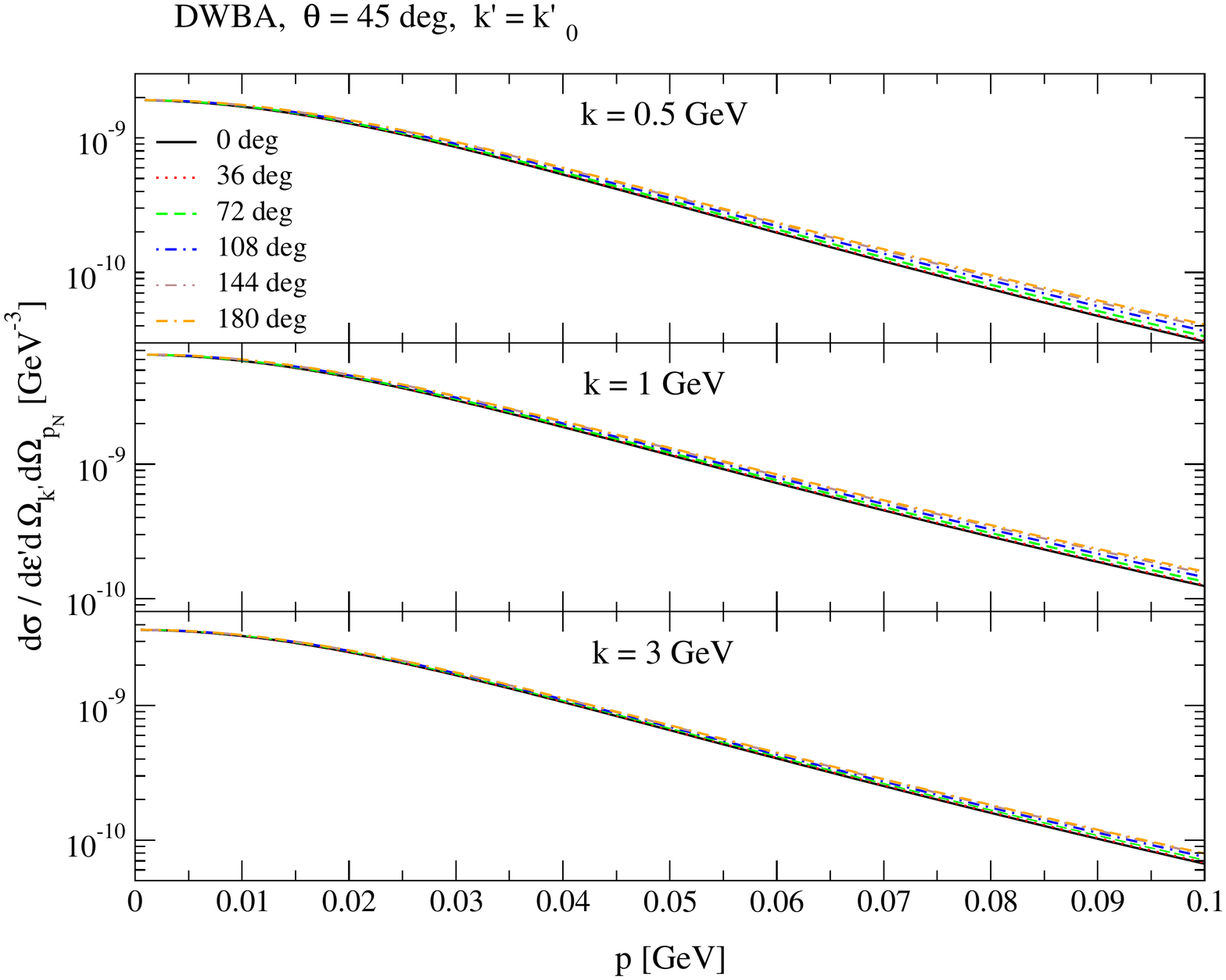}
\caption{(Color online) Semi-inclusive neutrino-deuteron cross sections for different nucleon emission angles $\phi$ within DWBA for three neutrino momenta $k=$ 0.5, 1, and 3 GeV, scattering angle $\theta=$ 45$^{\text{o}}$ and muon momentum $k'=k'_0$ (at the quasielastic peak).\label{semiinclusive_deuteron_comp_phi_DWBA}}
\end{center}
\end{figure}

\begin{figure}
\begin{center}
\includegraphics[width=0.8\textwidth]{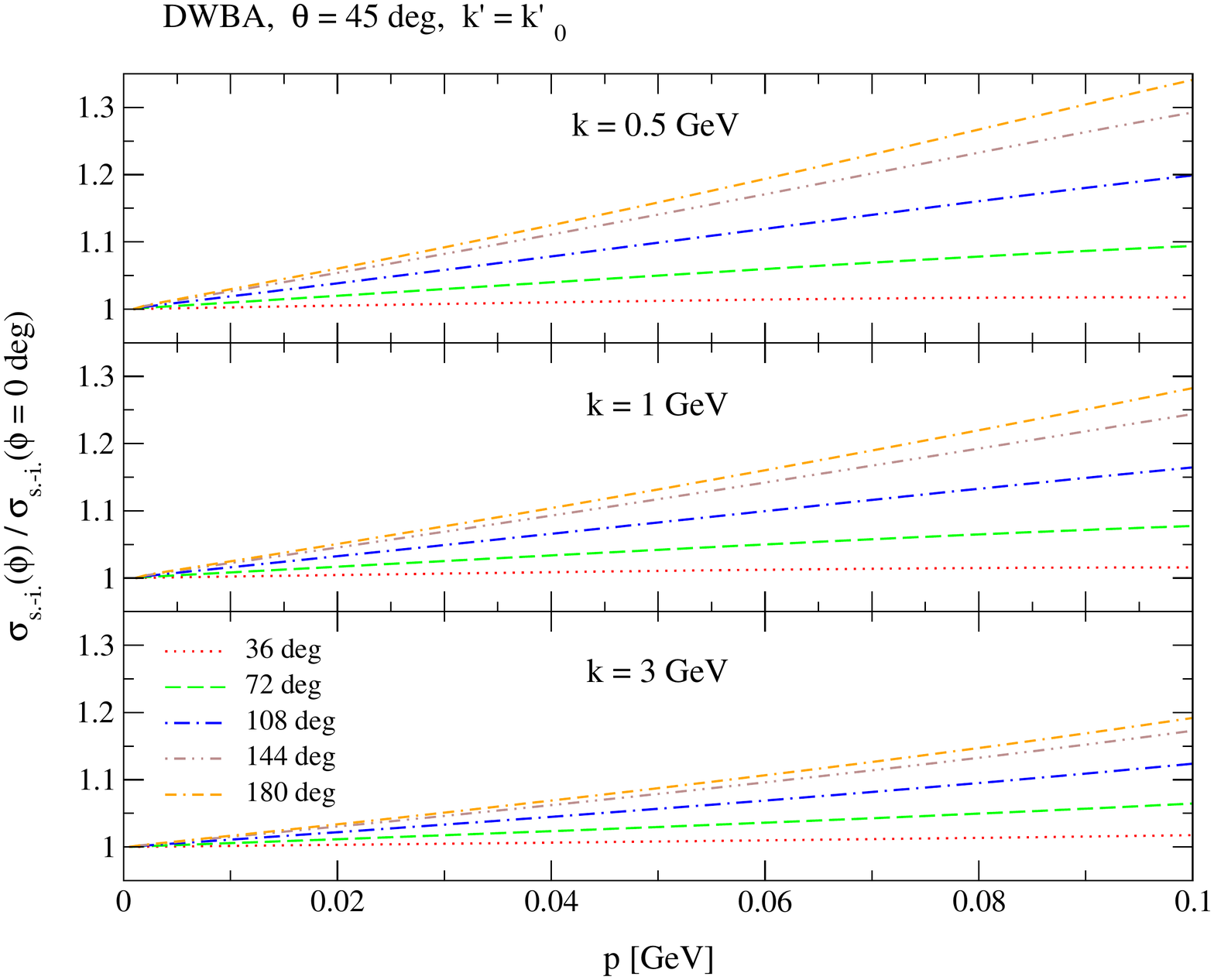}
\caption{(Color online) Ratios of the semi-inclusive neutrino-deuteron cross sections ($\sigma_{s.-i.}$) for different nucleon emission angles $\phi$ over the result for $\phi=$ 0$^{\text{o}}$, within DWBA for three neutrino momenta $k=$ 0.5, 1, and 3 GeV, scattering angle $\theta=$ 45$^{\text{o}}$ and muon momentum $k'=k'_0$ (at the quasielastic peak).\label{semiinclusive_deuteron_comp_phiratio_DWBA_k0}}
\end{center}
\end{figure}

\begin{figure}
\begin{center}
\includegraphics[width=0.75\textwidth]{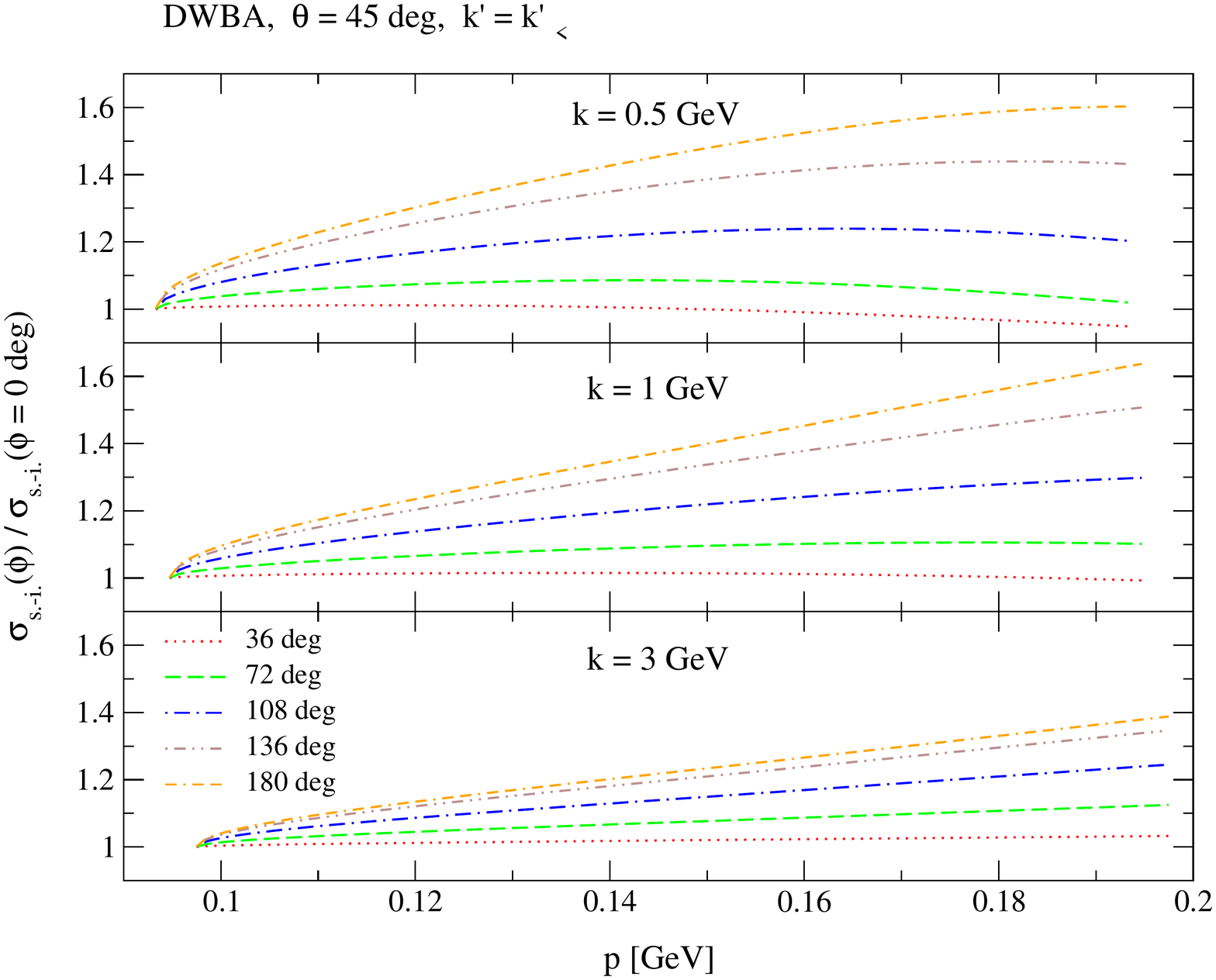}
\caption{(Color online) Same as in Fig. \ref{semiinclusive_deuteron_comp_phiratio_DWBA_k0} but for muon momentum $k'=k'_<$.
\label{semiinclusive_deuteron_comp_phiratio_DWBA_km}}
\end{center}
\end{figure}

In Fig. \ref{semiinclusive_deuteron_comp_phi_DWBA} we show semi-inclusive neutrino-deuteron cross sections for different nucleon emission angles $\phi$ for representative neutrino momenta and scattering angle in DWBA at the quasielastic peak ($k'=k'_0$). In order to clearly see the deviation of the cross sections due to the interference responses (the $\phi$-dependent terms) that contribute to the cross section in DWBA, we show in Fig. \ref{semiinclusive_deuteron_comp_phiratio_DWBA_k0} for the same kinematics at the peak and in Fig. \ref{semiinclusive_deuteron_comp_phiratio_DWBA_km} at $k'=k'_<$, the ratios of semi-inclusive neutrino-deuteron cross sections for different nucleon emission angles $\phi$ over the result for $\phi=$ 0$^{\text{o}}$. The deviations increase with increasing angle $\phi$, increasing missing momentum $p$ and decreasing incident momentum $k$, reaching up to a 40$\%$ at the quasielastic peak and up to a 60$\%$ at $k'=k'_<$. Upon integrating to get the inclusive no-pion cross section, all of these $\phi$-dependent contributions disappear; however, for semi-inclusive measurements there are clearly significant effects to be seen when comparing in-plane with out-of-plane detection of the final-state protons. 

\begin{figure}
\begin{center}
\includegraphics[width=0.8\textwidth]{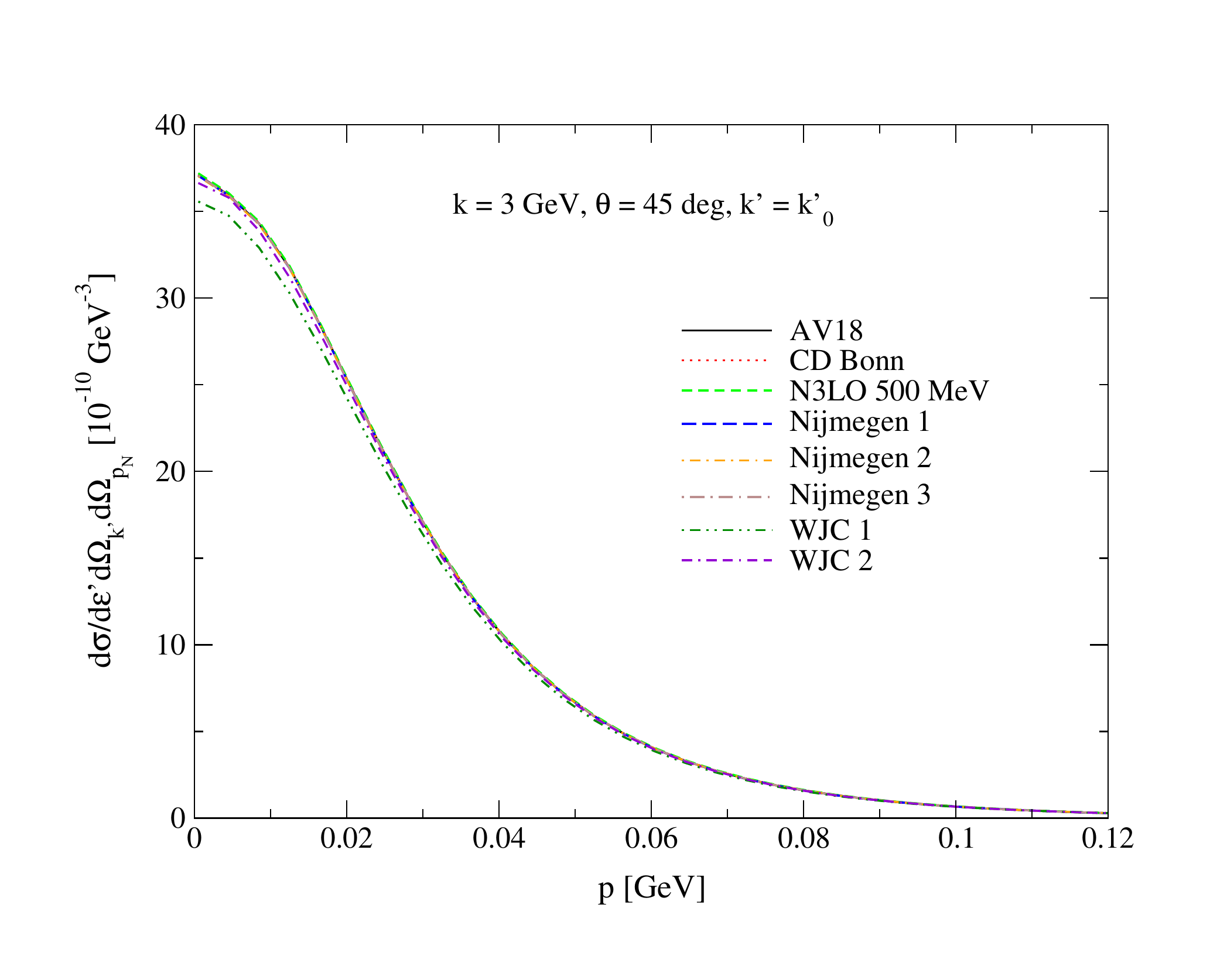}
\caption{(Color online) Semi-inclusive neutrino-deuteron cross sections using the deuteron wave functions AV18 \cite{Wiringa95}, CD Bonn \cite{Machleidt:2000ge}, N3L0500 \cite{Entem:2003ft}, Nijmegen 1, 2 and 3 \cite{Stoks:1994wp}, and WJC 1 and 2 \cite{Gross:2007jj} for $k=$ 3 GeV, $\theta=$ 45$^{\text{o}}$ and $k'=k'_0$.
\label{semiinclusive_deuteron_wavefunctions}}
\end{center}
\end{figure}

For completeness we have also studied the effect of initial-state interactions by using different deuteron wave functions. As an example we show in Fig. \ref{semiinclusive_deuteron_wavefunctions}, for representative kinematics, the CC$\nu$ deuteron disintegration semi-inclusive cross sections using different deuteron wave functions. For the kinematics shown, this set of wave functions yields results that differ by only a small amount. The largest deviation from the average is for the WJC 1 deuteron vertex function which differs for the rest at $p=0$ by less than 5$\%$. The interaction kernel in this case contains a small admixture of pseudoscalar pion exchange which produces a hard short-range force that causes cross section strength to be transfered from small to large $p$ resulting in an unusually small cross section at $p=0$.

\subsection{Effect of axial mass}

\begin{figure}
\begin{center}
\includegraphics[width=0.75\textwidth]{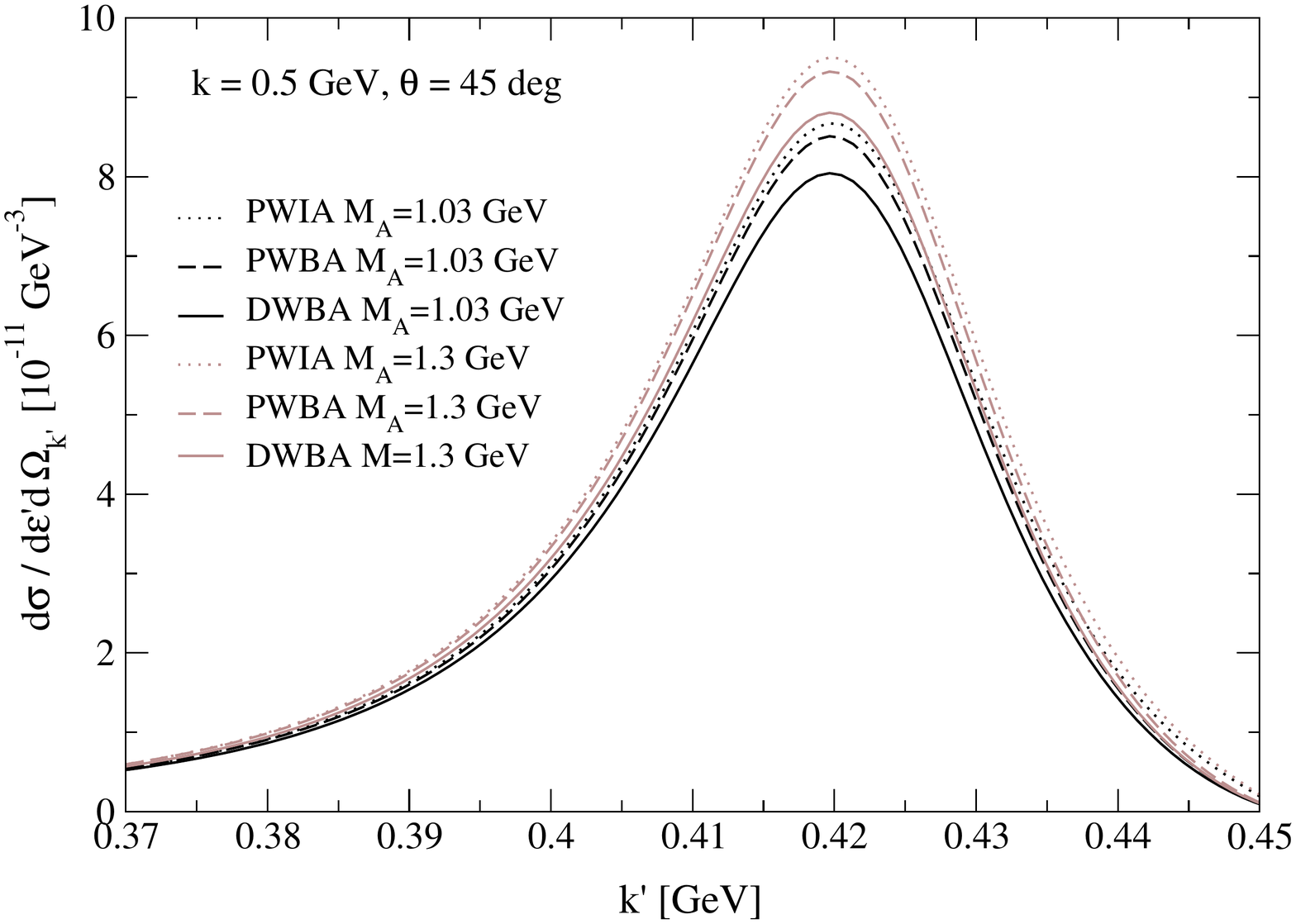}
\caption{(Color online) For axial masses $M_A=$ 1.03 GeV (dark curves) and $M_A=$ 1.3 GeV (lighter curves), CC$\nu$ deuteron disintegration inclusive no-pion cross section as a function of the outgoing muon momentum $k'$ for incident neutrino momentum $k=$ 0.5 GeV and scattering angle $\theta=$ 45$^{\text{o}}$. Dotted lines correspond to PWIA results, dashed lines to PWBA and solid lines to DWBA.
\label{inclusive_deuteron_axialmasscomp1}}
\end{center}
\end{figure}

The above results have all been obtained using an axial mass $M_A=$ 1.03 GeV in the parameterization of the axial form factor of Eq. (\ref{axial_ff}). We have computed the same inclusive no-pion cross sections using an increased value of the axial mass, $M_A=$ 1.3 GeV, and we have found that for small momentum transfers ($q \lesssim$ 150 MeV), the results hardly change as expected, since then the $Q^2$-dependence in the form factors is small and they are determined by their $Q^2 \rightarrow 0$ limit. For larger momentum transfers, however, the $Q^2$-dependence is now important and one sees that the increased axial mass gives noticeably larger inclusive no-pion cross sections. For momentum transfers between 150 MeV and 500 MeV, approximately, there is an overlap between the spread of results for different calculations (PWIA, PWBA and DWBA) obtained with the two values of the axial mass. Therefore, in that kinematic region the effect of exchange and/or distortion can be confused with a modified value of the axial mass, which calls for extra caution when interpreting experimental data.
For instance, inclusive no-pion neutrino-deuteron measurements at incident momentum $k=$ 0.5 GeV and scattering angle $\theta=$ 45$^{\text{o}}$ (corresponding to $q\approx$ 360 MeV at the peak), compatible with an axial mass $M_A=$ 1.03 GeV when interpreted through a PWIA calculation, would actually correspond to a larger axial mass, close to $M_A=$ 1.3 GeV, if they were (more correctly) interpreted using a DWBA calculation. This comparison is shown in Fig. \ref{inclusive_deuteron_axialmasscomp1} (results for $M_A=$ 1.03 GeV were also shown in the first plot of Fig. \ref{inclusive_deuteron2}).

\begin{figure}
\begin{minipage}[p]{0.48\linewidth}
\centering
\includegraphics[width=1.1\textwidth] {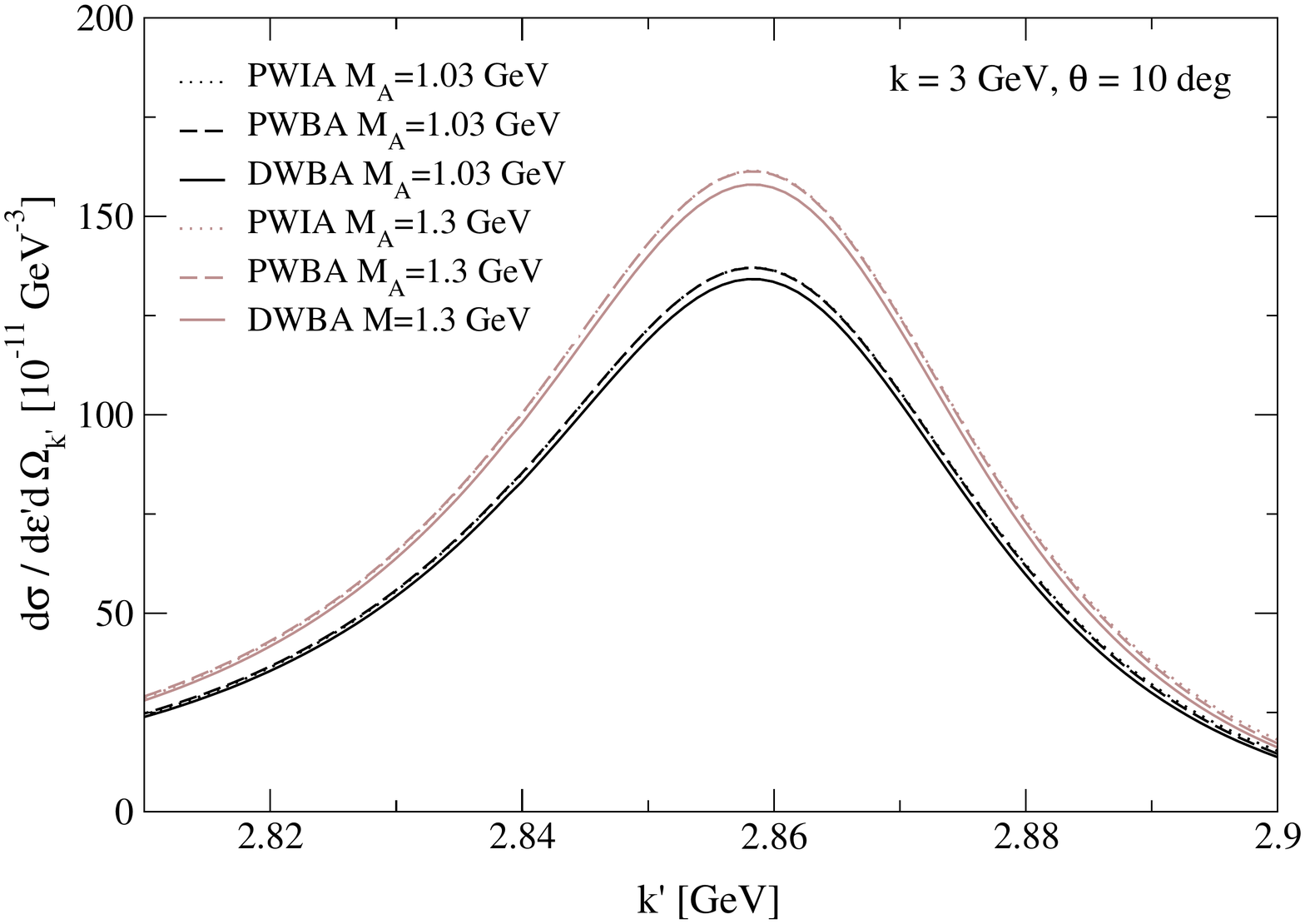}
\end{minipage}
\hspace{0.15in}
\begin{minipage}[p]{0.48\linewidth}
\centering
\includegraphics[width=1.1\textwidth] {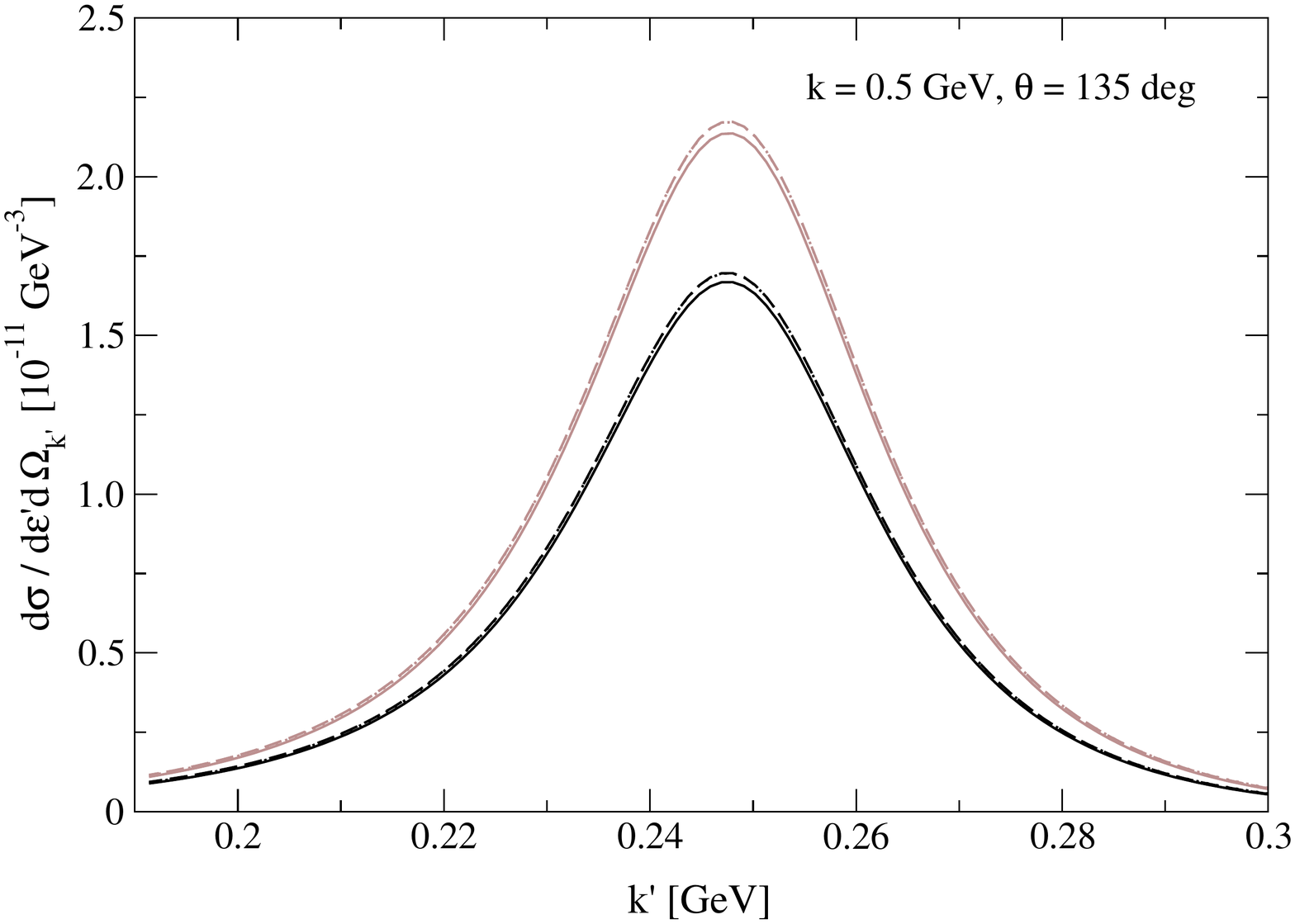}
\end{minipage}
\caption{(Color online) Same as in Fig. \ref{inclusive_deuteron_axialmasscomp1} but for $k=$ 3 GeV, $\theta=$ 10$^{\text{o}}$ (left), and for $k=$ 0.5 GeV, $\theta=$ 135$^{\text{o}}$ (right).
 \label{inclusive_deuteron_axialmasscomp2}}
\end{figure}

\begin{figure}
\begin{minipage}[p]{0.48\linewidth}
\centering
\includegraphics[width=1.1\textwidth] {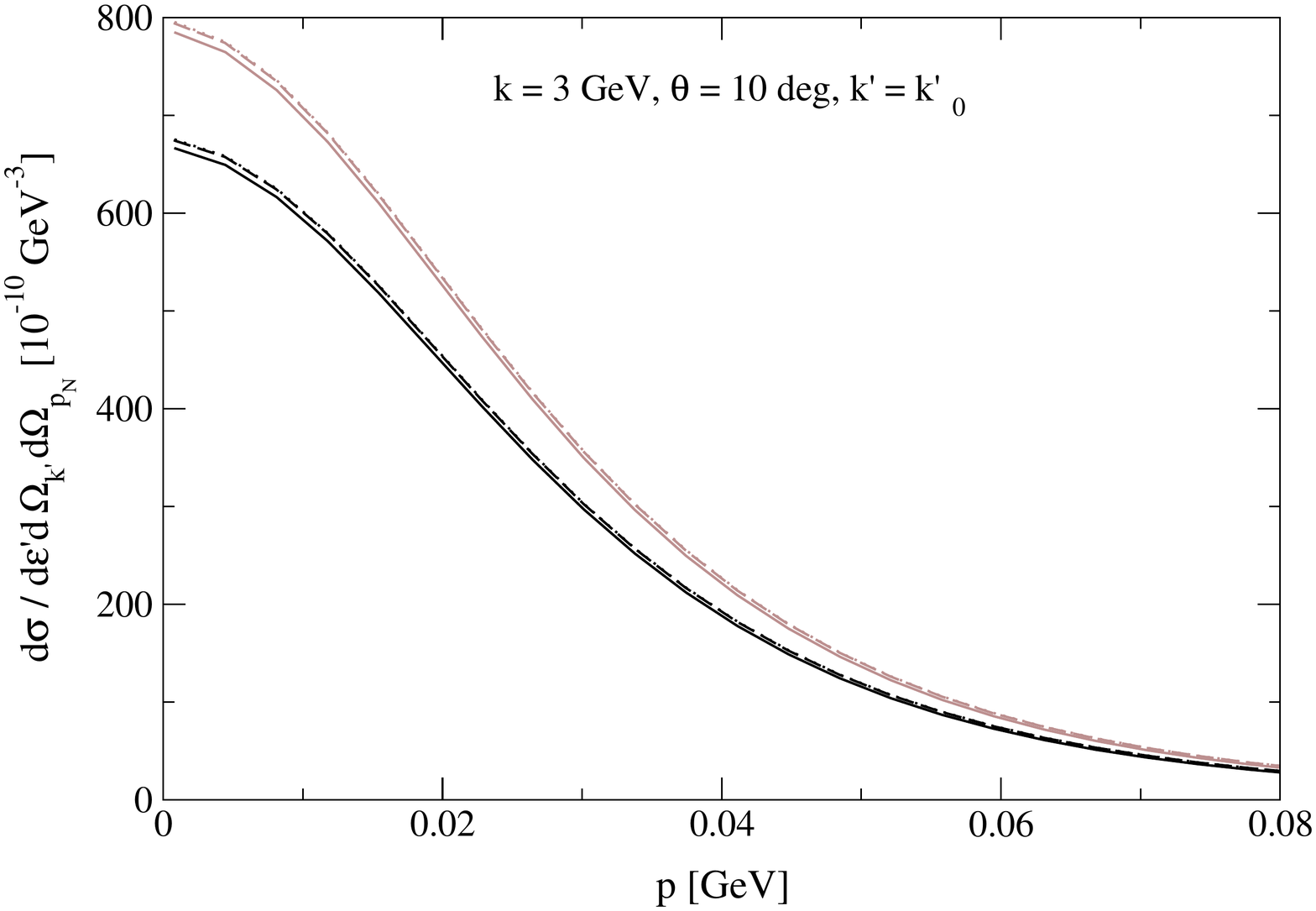}
\end{minipage}
\hspace{0.15in}
\begin{minipage}[p]{0.48\linewidth}
\centering
\includegraphics[width=1.1\textwidth] {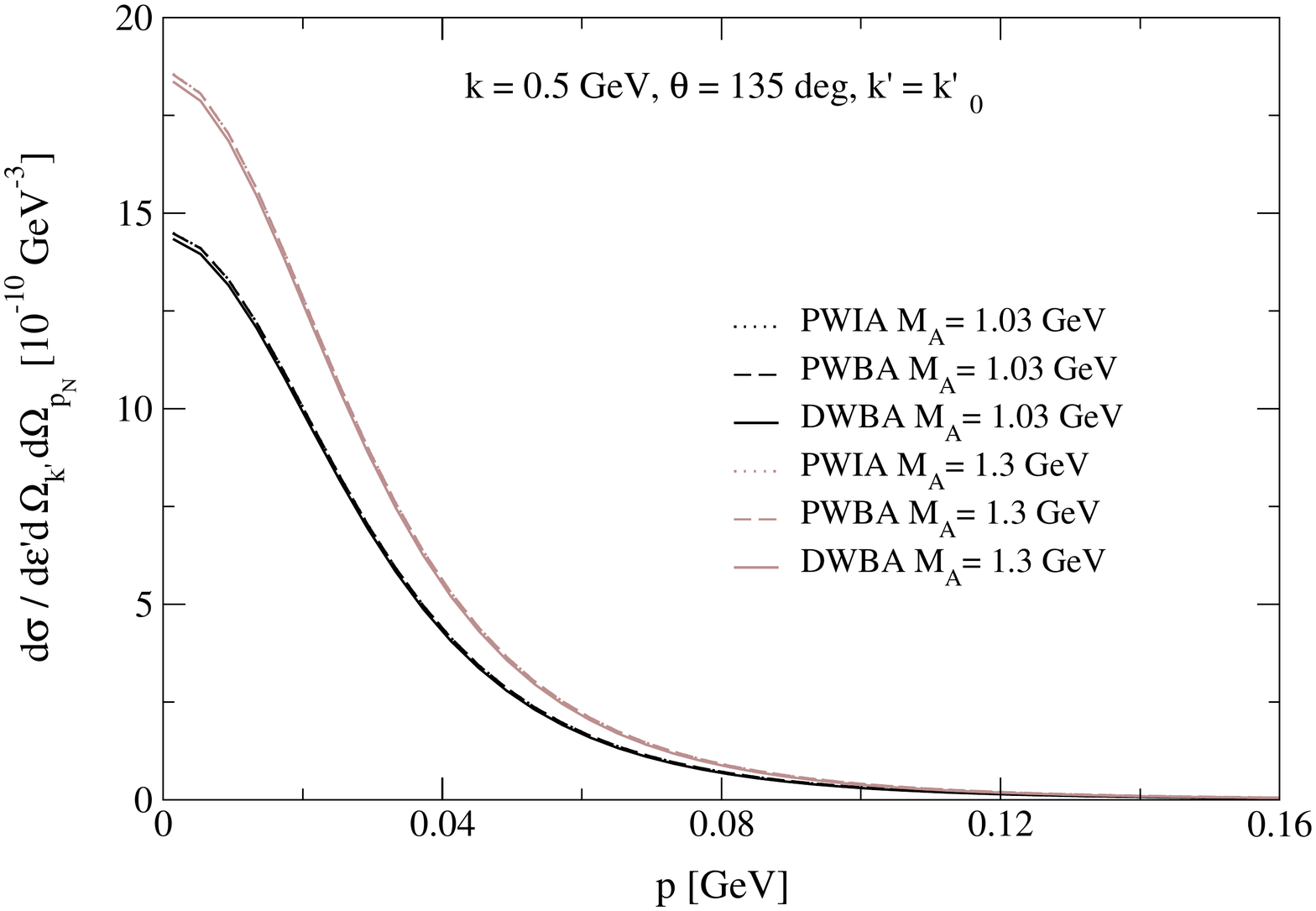}
\end{minipage}
\caption{(Color online) For axial masses $M_A=$ 1.03 GeV (dark curves) and $M_A=$ 1.3 GeV (lighter curves), CC$\nu$ deuteron disintegration semi-inclusive cross sections as a function of the missing momentum $p$ for the same kinematic conditions of Fig. \ref{inclusive_deuteron_axialmasscomp2}, namely  $k=$ 3 GeV, $\theta=$ 10$^{\text{o}}$ (left), and $k=$ 0.5 GeV, $\theta=$ 135$^{\text{o}}$ (right), and the muon momentum at the peak, $k'=k'_0$. Dotted lines correspond to PWIA results, dashed lines to PWBA and solid lines to DWBA.
 \label{semiinclusive_deuteron_axialmasscomp}}
\end{figure}

As said above, the three types of modeling give very similar results at momentum transfers between 500 and 1000 MeV. The negligible spread seen there might lead one to naively think that those kinematic conditions are well-suited to determining the axial mass from experiment, as long as data from other regions are ignored in the analysis to avoid confusion. We show in Fig. \ref{inclusive_deuteron_axialmasscomp2} examples of results in that region for $M_A=$ 1.03 GeV and for $M_A=$ 1.3 GeV: at incident momentum $k=$ 3 GeV and scattering angle $\theta=$ 10$^{\text{o}}$, corresponding to $q\approx$ 530 MeV at the peak (left), and at $k=$ 0.5 GeV and $\theta=$ 135$^{\text{o}}$, $q\approx$ 700 MeV (right).
However, even if the adjustment of the data to a given axial mass could be done in principle in that region, the corresponding momentum transfer is uncertain within a range related to the incident neutrino flux. As we showed above, only semi-inclusive measurements, not inclusive ones, allow one to determine the incoming neutrino momentum and therefore the momentum transfer of the process (within the experimental uncertainties).  Semi-inclusive measurements and calculations are thus required to precisely determine the axial mass. We show in Fig. \ref{semiinclusive_deuteron_axialmasscomp}, at the same kinematic conditions of the plots in Fig. \ref{inclusive_deuteron_axialmasscomp2} and at the quasielastic peak ($k'=k'_0$), the comparison  of semi-inclusive cross sections for different axial masses. Clearly, measuring such semi-inclusive cross sections has the potential of yielding an excellent way to determine the axial-vector form factor. The uncertainties from the vector form factors from electron scattering are relatively small, and, as we have seen, the uncertainties from modeling the initial- and final-state two-body problem is arguably the smallest to be found in the entire periodic table.

\section{Conclusions \label{conclusions}}

We have defined the kinematics of deuteron disintegration by neutrino scattering, $\nu_{\mu} + {}^2$H$\rightarrow \mu^- + p + p$, and analyzed different detection scenarios in terms of both theoretically natural and experimentally practical kinematic variables. The deuteron as a target has the peculiarity that a {\em semi-inclusive} (coincidence) measurement of any two of the three particles in the final-state is actually {\em exclusive}, since the final state, consisting of a charged lepton and two protons (below the pion production threshold), has no other open channel, in contrast to the usual situation for complex nuclei where both missing momentum and missing energy dependences occur. One obvious conclusion is that semi-inclusive studies of this reaction hold promise for determining the incident neutrino energy using kinematics alone.

In this study we have presented results for both inclusive no-pion and semi-inclusive CC$\nu$ cross sections using for the underlying dynamics a relativistic model of the deuteron and final-state $pp$ structure that involves an approximation to the Bethe-Salpeter equation.  We note that such an approach is completely covariant: the kinematics, the initial-state deuteron, the current operators and the final-state $pp$ system are all fully relativistic, in contrast to much of the modeling of CC$\nu$ reactions in general, where approximations are often employed (for a discussion of the role played by relativity for such reactions see \cite{Garvey:2014exa}). We have provided inclusive no-pion and semi-inclusive CC$\nu$ deuteron disintegration cross sections for a variety of kinematical conditions within PWIA, PWBA ({\it i.e.}, introducing final-state hadronic exchange terms) and DWBA  ({\it i.e.}, using final-state hadronic interactions including exchange). The calculation of the two-body dynamics with the model used in this work, particularly when including exchange terms and final-state interactions (DWBA), is computationally demanding and therefore not well suited as direct input to  neutrino event generators. However, we anticipate being able to provide simple parameterizations of the cross section in work that is currently in progress, and these will soon be made available for event simulations.

At small and large momentum transfers ($q\lesssim 500$ and $q\gtrsim 1000$ MeV) the approximation used for the interaction (PWIA, PWBA, DWBA) plays a significant role. Clearly antisymmetrization (going from PWIA to PWBA) is seen to be important for such kinematics, with a somewhat smaller effect seen to arise from final-state interactions (going to the full DWBA). On the other hand, initial-state variations using state-of-the-art deuteron wave functions are observed to be minor. In contrast, for intermediate values of $q$, the three types of approximation lead to essential a universal result, showing very small sensitivity to both initial- and final-state physics. Thus, with an appropriate choice of kinematics, namely where the $Q^2$-dependence of the axial-vector form factor plays a minor role (see below), results of semi-inclusive CC$\nu$ disintegration of deuterium could not only provide the incident neutrino energy, as stated above, but could also yield the incident neutrino flux with relatively minor uncertainties from the modeling.

Finally, in addition to the motivations for studying the $\nu_{\mu} + {}^2$H$\rightarrow \mu^- + p + p$ reaction given above, it can also serve as a way to improve our knowledge of the $Q^2$-dependence of the isovector, axial-vector form factor. We have seen that the impact of changing the axial mass $M_A$ used in a dipole parameterization of this form factor is significant in regions where the dependence on the modeling of the two-body problem is weak. The $M_A$-dependence increases for increasing momentum transfer, being negligible at $q\lesssim 150$ MeV. We have shown that there is a range of intermediate momentum transfers (150 $\lesssim q \lesssim$ 500 MeV) where the spreads of the curves for different approximations (PWIA, PWBA, DWBA) computed with axial masses $M_A\sim 1$ and $M_A\sim 1.3$ GeV overlap, leading to a potential confusion in the extraction of the axial mass from experimental data unless the full model is employed. We close by noting that at some level inclusive measurements always suffer from their inability to be performed at a specific value of $q$, whereas the ability to determine all kinematic variables using semi-inclusive measurements, including the momentum transfer, provides a unique tool for studies of this type.

\begin{acknowledgments}
O. M. acknowledges support from a Marie Curie International Outgoing Fellowship within the 7th Framework Programme of the European Union. Also supported in part by the Office of Nuclear Physics of the US Department of Energy under Grant Contract DE-FG02-94ER40818 (T. W. D.), and by the US Department of Energy under Contract No. DE-AC05-06OR23177 and the U.S. Department of Energy cooperative research agreement DE-AC05-84ER40150 (J. W. V. O.). 
\end{acknowledgments}

\bibliography{neutrino}

\end{document}